\documentstyle[prc,aps,epsfig]{revtex}

\begin{document}

\draft

\title{Electromagnetic Dissociation of $^{8}$B and the Rate of the
$^{7}$Be(\textit{p},$\gamma$)$^{8}$B Reaction in the Sun}

\author {B.~Davids$^{1,2}$\cite{kvi}, Sam~M.~Austin$^{1,2}$, D.~Bazin$^{1}$,
H.~Esbensen$^{3}$, B. M.~Sherrill$^{1,2}$, I.~J.~Thompson$^{4}$, and
J.~A.~Tostevin$^{4}$}

\address {$^{1}$ National Superconducting Cyclotron Laboratory, Michigan State
University, East Lansing, Michigan 48824\\$^{2}$ Department of Physics and
Astronomy, Michigan State University, East Lansing, Michigan 48824\\$^{3}$
Physics Division, Argonne National Laboratory, Argonne, Illinois 60439\\$^{4}$
Department of Physics, School of Physics and Chemistry, University of Surrey,
Guildford, United Kingdom}

\date{\today}

\maketitle

\begin{abstract}In an effort to better determine the
$^{7}$Be(\textit{p},$\gamma$)$^{8}$B reaction rate, we have performed inclusive
and exclusive measurements of the Coulomb dissociation of $^{8}$B. The former
was a study of longitudinal momentum distributions of $^{7}$Be fragments emitted
in the Coulomb breakup of intermediate energy $^{8}$B beams on Pb and Ag
targets. Analysis of these data yielded the $E$2 contribution to the breakup
cross section. In the exclusive measurement, we determined the cross section for
the Coulomb breakup of $^8$B on Pb at low relative energies in order to infer
the astrophysical $S$ factor for the $^{7}$Be(\textit{p},$\gamma$)$^{8}$B
reaction. Interpreting the measurements with 1st-order perturbation theory, we
obtained $S_{E2}/S_{E1}$ = 4.7 $^{+ 2.0}_{- 1.3} \times 10^{-4}$ at $E_{rel}$ =
0.6 MeV, and $S_{17}(0)$ = 17.8 $^{+ 1.4}_{- 1.2}$ eV~b. Semiclassical 1st-order
perturbation theory and fully quantum mechanical continuum-discretized coupled
channels analyses yield nearly identical results for the E1 strength relevant to
solar neutrino flux calculations, suggesting that theoretical reaction mechanism
uncertainties need not limit the precision of Coulomb breakup determinations of
the $^{7}$Be(\textit{p},$\gamma$)$^{8}$B $S$ factor. A recommended value of
$S_{17}(0)$ based on a weighted average of this and other measurements is
presented. \end{abstract}


\section{Introduction}

The flux of neutrinos emanating from the solar interior consists predominantly
of low energy electron neutrinos from the $\textit{p} + \textit{p} \rightarrow
\textit{d} + e^{+} + \nu_{e}$ reaction \cite{bahcall}. The higher energy
$^{8}$B neutrinos, though they constitute less than 10$^{-4}$ of the total
solar neutrino flux, are the best studied. Their flux, direction, and energy
spectrum have been measured in large chlorine radiochemical and water
\^{C}erenkov detectors. Less than half the number of $^{8}$B neutrinos expected
on the basis of standard solar models and standard electroweak particle physics
has been observed in terrestrial detectors \cite{superk}, a situation that has
come to be known as the $^{8}$B solar neutrino problem. This discrepancy
between theory and experiment appears to be resolved best by invoking
oscillations of $\nu_{e}$ into other neutrino flavors. By measuring the ratio
of charged current to neutral current interactions, the heavy water detector
SNO will stringently test neutrino oscillation hypotheses. In order to
calculate the theoretical solar neutrino flux and interpret the results of
measurements at SNO and other neutrino detectors, the rate of the radiative
capture reaction that produces $^{8}$B in the Sun,
$^{7}$Be(\textit{p},$\gamma$)$^{8}$B, must be known to a precision of 5\%
\cite{bahcall}. Thus astrophysics, nuclear physics, and particle physics meet
in addressing the solar neutrino problem.

The cross section for the $^{7}$Be(\textit{p},$\gamma$)$^{8}$B reaction has
been measured directly in several experiments
\cite{kavanagh60,parker,kavanagh69,vaughn,wiezorek,filippone,hass,hammache}.
Although the shape of the excitation function is fairly well determined, there
is a large spread in the absolute normalizations of these measurements. The
cross section must be known at the very low relative energies ($\sim$ 20 keV)
relevant to $^{8}$B production in the Sun, but these low energies are
experimentally inaccessible because the high Coulomb barrier causes the cross
section to plummet with decreasing energy. The strategy adopted is to measure
the cross section at the lowest possible energy, and then extrapolate downward
using theory. In order to extrapolate to low energies reliably, the dominant
energy dependences in the cross section can be factored out, leaving a quantity
known as the astrophysical $S$ factor, which varies much more slowly with
energy. The $S$ factor is defined by $S(E)~=~E~\sigma(E)~exp[2 \pi Z_{1} Z_{2}
e^{2} / (\hbar v)]$, where the $Z_{i}$ are the charges, $v$ the relative
velocity, and E the center-of-mass energy of the nuclei involved.
Conventionally, the value of the $S$ factor for the
$^{7}$Be(\textit{p},$\gamma$)$^{8}$B reaction, $S_{17}$, is extrapolated from
the data at accessible energies to zero energy.

In light of the disagreements among the radiative capture measurements of the
$^{7}$Be(\textit{p},$\gamma$)$^{8}$B cross section, and the fact that direct
measurements at very low energies are impractical, indirect techniques have been
developed to infer this radiative capture cross section. Such techniques are
subject to different systematic uncertainties. For photons of a given
multipolarity, the detailed balance theorem relates the cross section for
radiative capture to that for the corresponding inverse reaction,
photodissociation. In the case of the $^{7}$Be(\textit{p},$\gamma$)$^{8}$B
reaction, the $^{8}$B nucleus is radioactive with a half life of 770 ms and is
not a viable photodissociation target. However, when an energetic beam of
$^{8}$B nuclei passes through a heavy target, the time-dependent electromagnetic
field of the high Z target nuclei acts as a source of virtual photons capable of
dissociating the incident $^{8}$B projectiles into $^{7}$Be + $\textit{p}$
\cite{baur86}. This process, known as Coulomb dissociation, is Coulomb
excitation to the continuum. The semiclassical formalism of Coulomb excitation
has been extended to Coulomb dissociation at intermediate and high energies
\cite{winther,bertulani88,baur96}. The advantages of high energy Coulomb
dissociation over direct radiative capture measurements include thicker targets
and larger cross sections, and thereby the possibility of reaching lower
relative energies with an appreciable yield.

Coulomb dissociation has been used to infer $S_{17}(0)$
\cite{motobayashi,kikuchi98,iwasa}, but the method is not without complications.
First, several electromagnetic multipoles contribute in Coulomb dissociation,
e.g., $E1$, $E2$, and $M1$, while the radiative capture reaction is mainly
driven by a single electromagnetic multipole transition at solar energies, $E1$.
Second, even though the electromagnetic interaction dominates, the effects of
nuclear absorption and diffraction must be considered. Finally, one must
consider the effects of higher-order electromagnetic transitions that can
destroy the simple correspondence between radiative capture and Coulomb
dissociation. An important experimental challenge is to identify and understand
these complications in order to firmly establish Coulomb dissociation as a
viable alternative to direct radiative capture measurements. In this paper, we
will show that these complications can be dealt with in a satisfactory manner by
a judicious choice of the experimental conditions and by applying tested nuclear
structure and reaction theories.

The first challenge, that of disentangling the contributions of different
electromagnetic multipoles to extract one in particular, can be met by carefully
studying the angular distribution of the breakup fragments. In the Coulomb
dissociation of intermediate energy $^{8}$B, $E2$ is the principal unwanted
electromagnetic multipole. By carrying out an inclusive measurement of the
$^{7}$Be fragments emitted in the Coulomb dissociation of 44 and 81 MeV/nucleon
$^{8}$B, we determined the $E2$ contribution to the breakup cross section.

Having measured the $E2$ strength in the Coulomb breakup, we were in a position
to study the breakup energy spectrum, $d\sigma/dE_{rel}$, in order to determine
the $E1$ strength at low relative energies and thereby infer $S_{17}(0)$. This
was done in an exclusive measurement of the elastic breakup of an 83 MeV/nucleon
$^{8}$B beam. In the analysis of this experiment, we used the $E2$ strength
determined in the inclusive measurement, and investigated the influence of
nuclear and higher-order electromagnetic processes. We discovered that these
complications could be minimized, and the theoretical uncertainties made small
enough that our Coulomb breakup measurement is of comparable precision to the
direct radiative capture measurements. In this paper, we shall describe the
inclusive and exclusive measurements, and interpret them using both 1st-order
perturbation theory and a continuum-discretized coupled channels approach.
Finally we will compare the inferred value of $S_{17}(0)$ with recent direct and
indirect measurements.

\section{Experimental Procedures}

\subsection{Inclusive Measurement}

We bombarded a 1.9 g cm$^{-2}$ Be production target with 100 and 125 MeV/nucleon
$^{12}$C beams from the K1200 cyclotron at the National Superconducting
Cyclotron Laboratory (NSCL). Typical $^{12}$C beam intensities were 10-50 pnA.
Fragmentation reactions in the Be target yielded secondary beams of 44 and 81
MeV/nucleon $^{8}$B, after magnetic analysis in the A1200 fragment separator
\cite{A1200}. A 200 mg cm$^{-2}$ (CH$_{2}$)$_n$ achromatic energy degrader aided
in the purification of the secondary beams. Slits limited the momentum spread of
the secondary beams to $\pm$ 0.25\%. A 17 mg cm$^{-2}$ plastic scintillator just
downstream of the A1200 focal plane provided time-of-flight and secondary beam
intensity information. Typical $^{8}$B beam intensities ranged from (4-20)
$\times$ 10$^{3}$ s$^{-1}$. Table\ \ref{table1} shows the total number of
$^{8}$B nuclei that struck each target.

\begin{figure}\epsfig{file=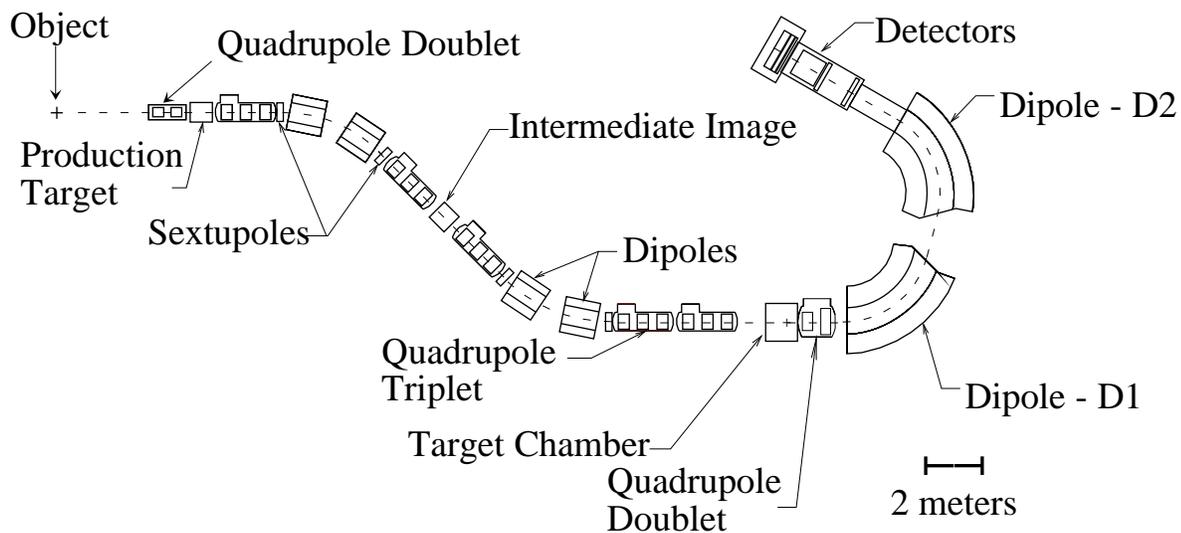} \caption{Schematic view of the S800
spectrometer at the NSCL.} \label{S800} \end{figure}

\begin{figure}\epsfig{file=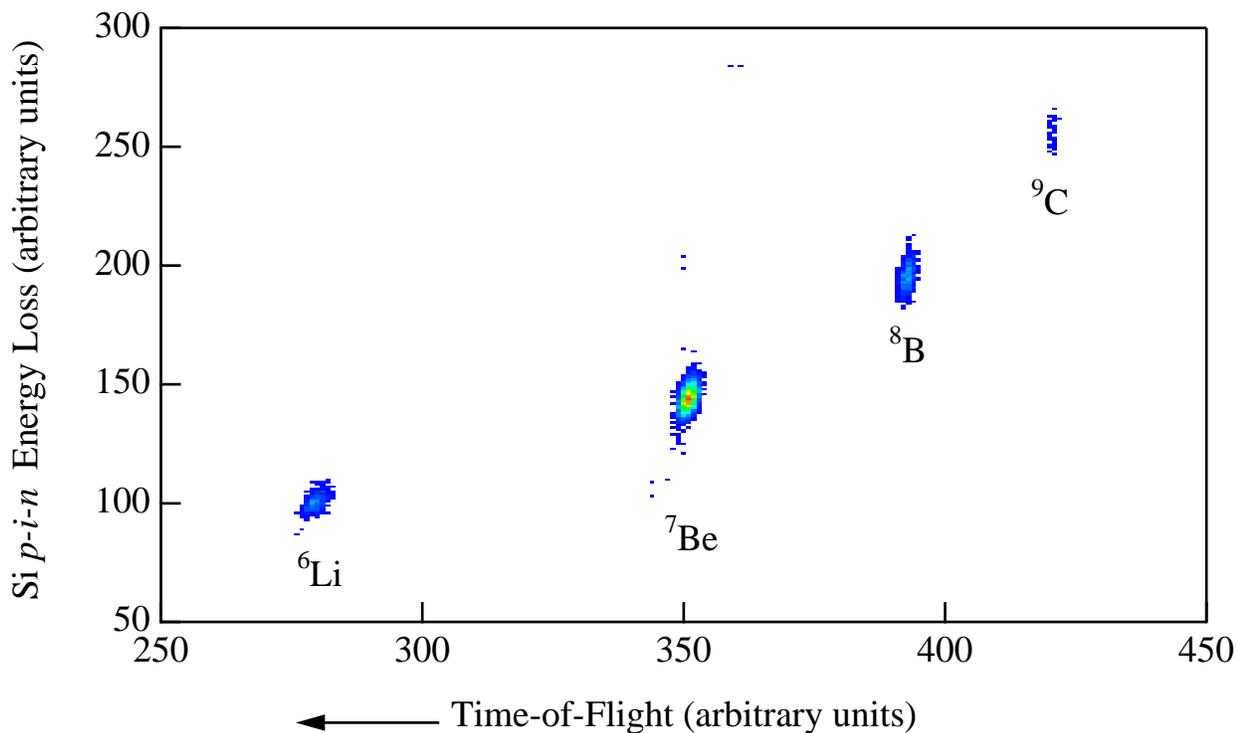} \caption{Typical Si $\textit{p-i-n}$
diode versus time-of-flight spectrum illustrating the secondary beam
composition.} \label{pintof} \end{figure}

The $^{8}$B beams were transported through a second beam analysis line to the
target position of the S800 spectrometer, shown schematically in Fig.\
\ref{S800}. This analysis line dispersed the secondary beams according to their
momenta, resulting in a 5 cm $\times$ 1 cm beam spot on the targets. A ladder
held 27 mg cm$^{-2}$ Ag and 28 mg cm$^{-2}$ Pb targets. A 300 $\mu$m Si
$\textit{p-i-n}$ diode detector mounted on a ladder 18 cm upstream of the
targets was intermittently raised into the path of the beam. Energy loss
signals from this detector, in conjunction with timing signals from the plastic
scintillator at the exit of the A1200, yielded both the transmission and
composition of the secondary beams. Times-of-flight were measured for the
$\sim$ 70 m flight path between the scintillator at the exit of the A1200 and
the S800 focal plane. Fig.\ \ref{pintof} shows a typical plot of the signals in
the $\textit{p-i-n}$ diode detector versus time-of-flight.

The secondary beams were not monoisotopic; $^{7}$Be was the principal
contaminant, and was 5-8 times more intense than the $^{8}$B component of the
beam. Two other nuclei, $^{6}$Li and $^{9}$C, were also present in the beam. As
the velocities of these contaminants differed substantially from that of the
$^{8}$B ions, their different times-of-flight provided reliable particle
identification.

We used the S800 spectrometer \cite{jac} to detect the $^{7}$Be fragments
emitted in the Coulomb dissociation of $^{8}$B nuclei on the Ag and Pb targets.
The spectrometer was set at 0$^{\circ}$, and was operated in a
dispersion-matched energy loss mode so that the 0.5\% spread in the momentum of
the $^{8}$B beams did not limit the final momentum resolution, which was
dominated by differential energy loss in the target. The large angular
acceptance (20 msr) and momentum acceptance (6\%) of the S800 allowed us to
capture essentially the entire momentum distribution at a single magnetic field
setting.

\begin{figure}\epsfig{file=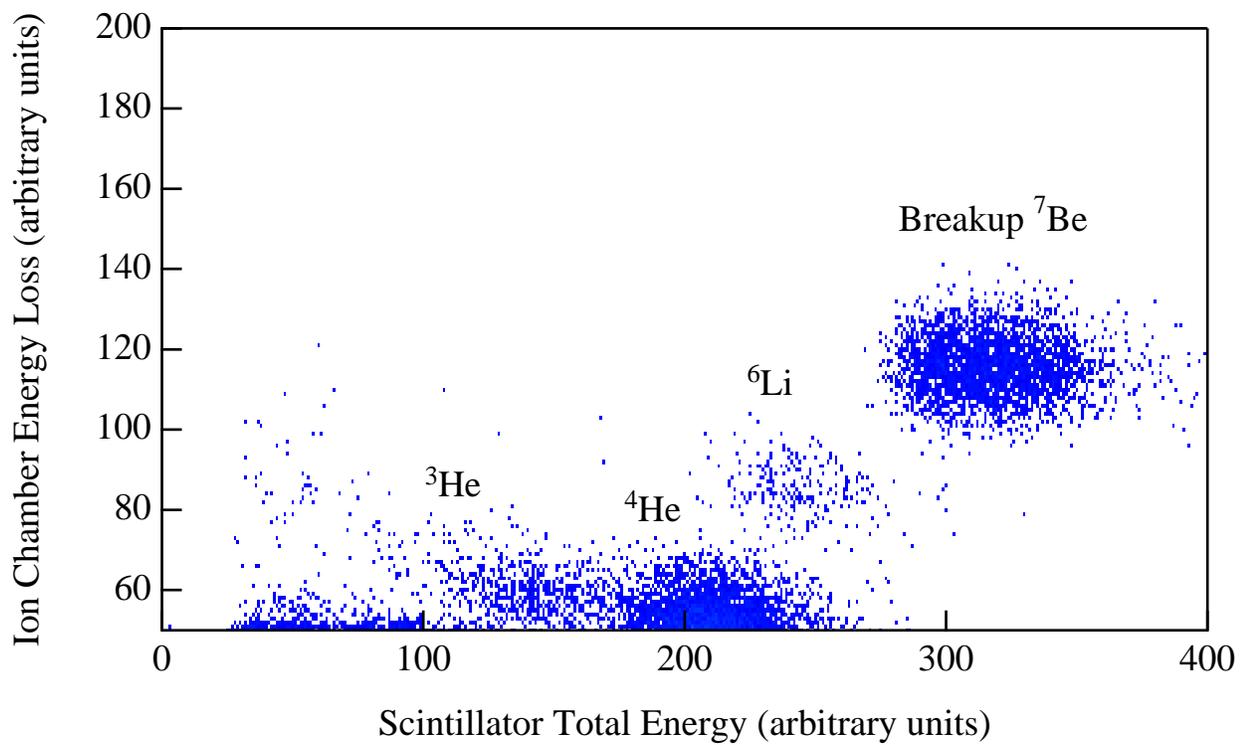} \caption{Typical ionization chamber
energy loss versus stopping scintillator total energy spectrum.} \label{S800pid}
\end{figure}

The standard complement of detectors at the focal plane of the S800 spectrometer
comprises two position-sensitive cathode readout drift chambers (CRDCs), a 41 cm
deep, 16 segment ionization chamber, and three plastic scintillators. The CRDCs
are separated by 1 meter to give good angular resolution. Ref.\ \cite{yurkon}
describes these detectors in detail. The ionization chamber recorded the energy
losses, and the first, 5 cm thick scintillator measured the total energies of
particles reaching the focal plane. This information was sufficient to identify
the $^7$Be breakup fragments unambiguously, as illustrated in Fig.\
\ref{S800pid}; the time-of-flight data provided a check. As the nuclei of
interest were stopped in the first scintillator, the other two were not used.
The particle identification was confirmed through comparisons with calibration
beams of $^{7}$Be that had the same magnetic rigidity as the $^{8}$B beams. The
higher magnetic rigidity of the detected $^{7}$Be fragments compared to the
$^{8}$B beams made the focal plane particle identification particularly clean.

The positions and angles of the $^{7}$Be fragments were measured in the CRDCs.
The position resolution obtained was approximately 0.3 mm (1$\sigma$), yielding
an intrinsic angular resolution of about 2 mrad. We employed the ion optics code
COSY INFINITY \cite{cosy1} to reconstruct the trajectories of the $^{7}$Be
fragments from their measured positions in the CRDCs, and the magnetic fields of
the spectrometer, which were continuously monitored by nuclear magnetic
resonance probes throughout the experiment. We calculated the $^{7}$Be lab
momenta and scattering angles on an event-by-event basis, allowing
reconstruction of the longitudinal momentum distributions.

Corrections to the momentum distributions were made for two different effects.
First, the overall efficiency of the CRDCs was less than unity due to a high
threshold on the anode wire constant fraction discriminator. This was a small
correction in the case of the low energy beam ($<$ 3\%), but larger for the high
energy beam ($<$ 15\%). The second was a momentum-dependent correction for the
angular acceptance of the S800, which was important for events having large
deviations from the central momentum and large projections of the scattering
angle in the dispersive direction of the spectrometer. These corrections
affected only the tails of the measured momentum distributions, and amounted to
less than 5\% of the measured cross sections, even for the largest scattering
angles. Corrections were made on the basis of the data themselves by
observations of the acceptance limits. Uncertainties equal to half the size of
the corrections were assigned to the data points in the momentum distributions
that required correction. During some runs, the magnetic field of the
spectrometer was varied to move the center of the distribution away from the
center of the focal plane detectors in order to measure the tails of the
momentum distributions precisely. The final momentum distributions represent the
sums of measurements made at several different magnetic field settings.

\subsection{Exclusive Measurement}

\begin{figure}\epsfig{file=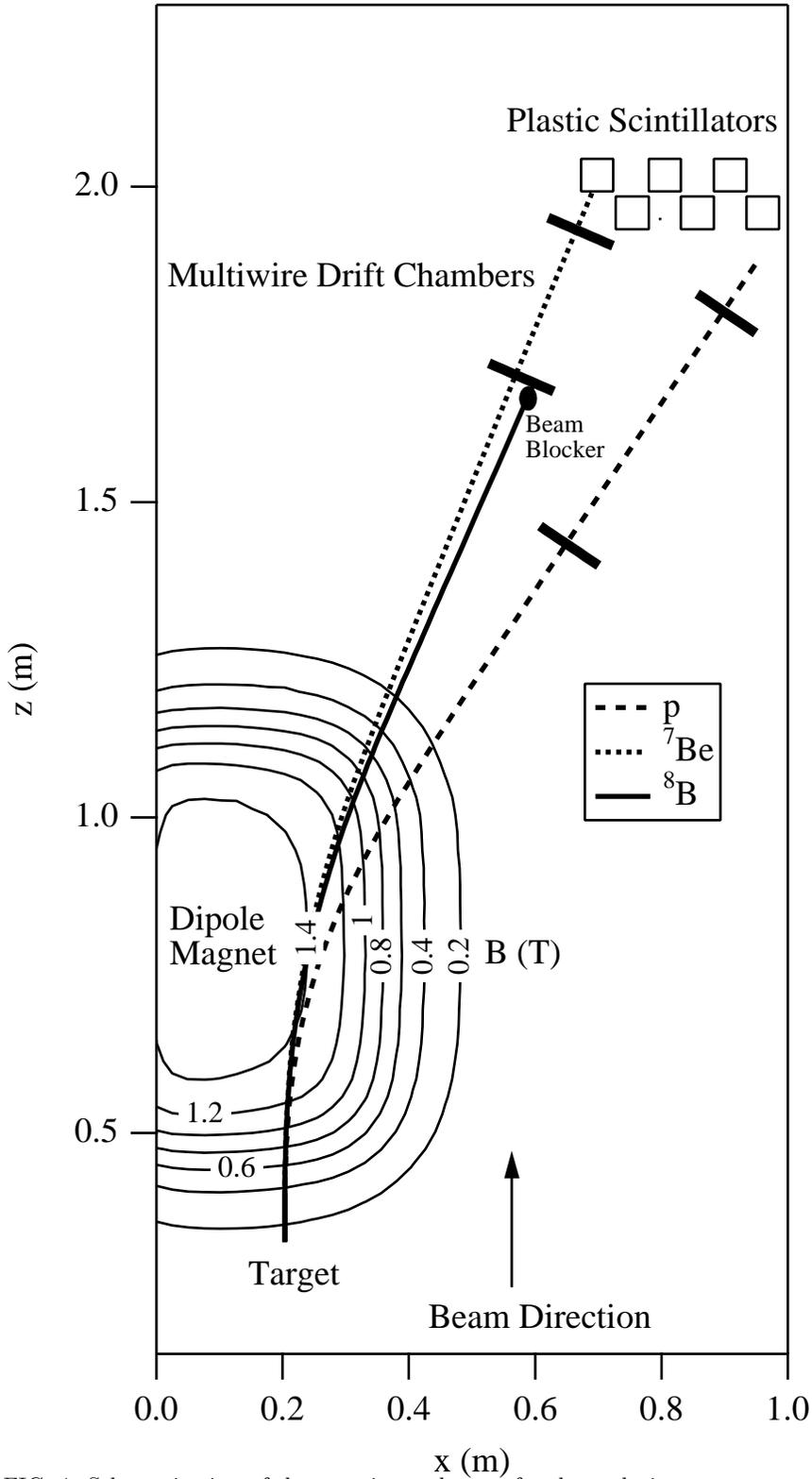} \caption{Schematic view of the experimental
setup for the exclusive measurement showing the detectors, typical trajectories,
and contours of constant magnetic field produced by the dipole magnet.}
\label{pos} \end{figure}

The 83 MeV/nucleon $^{8}$B beam used in the exclusive measurement was produced
with a 125 MeV/nucleon primary $^{12}$C beam in the same manner described above.
Typical $^8$B beam intensities were 10$^4$ s$^{-1}$, and the momentum spread in
the beam was limited to $\pm$ 0.25\% by slits in the A1200 fragment separator. A
total of $4 \times 10^9$ $^8$B nuclei were incident on the target. A thin
plastic scintillator was placed at the exit of the A1200 fragment separator for
beam intensity, transmission, and time-of-flight measurements. The $^{8}$B
nuclei were dissociated in a 47 mg cm$^{-2}$ Pb target located in front of a
room temperature 1.5 T dipole magnet. Four position-sensitive multiwire drift
chambers (MWDCs) \cite{mwdc} recorded the positions of the $^{7}$Be and
$\textit{p}$ fragments produced in the breakup after they passed through the
magnetic field. Two MWDCs measured each breakup fragment, allowing the
determination of both position and angle. A 16 element array of 4 cm thick
plastic scintillator bars was placed behind the MWDCs. A 25 mm $\times$ 60 mm
stainless steel plate located directly in front of the first $^{7}$Be MWDC
absorbed nearly all of the direct beam. The composition of the secondary beam in
the exclusive measurement was roughly 20\% $^6$Li, 55\% $^7$Be, 20\% $^8$B, and
5\% $^9$C. Fig.\ \ref{pos} shows a schematic drawing of the experimental setup.

The multiwire drift chambers used in this experiment have active areas of 112 mm
$\times$ 112 mm, and use delay-line readout to measure the positions of particle
tracks. The chambers were filled with P30 (70\% argon, 30\% methane) at a
pressure of 700 torr. Each MWDC has two orthogonal wire planes, providing both x
and y positions. Drift time information is used to interpolate between anode
wires using current pulses induced on the cathode field-shaping wires. The
principles of operation of these detectors are described in Ref. \cite{mwdc}.
Position resolutions of 0.4 mm (1$\sigma$) were obtained for protons and
$^{7}$Be fragments.

\begin{figure}\epsfig{file=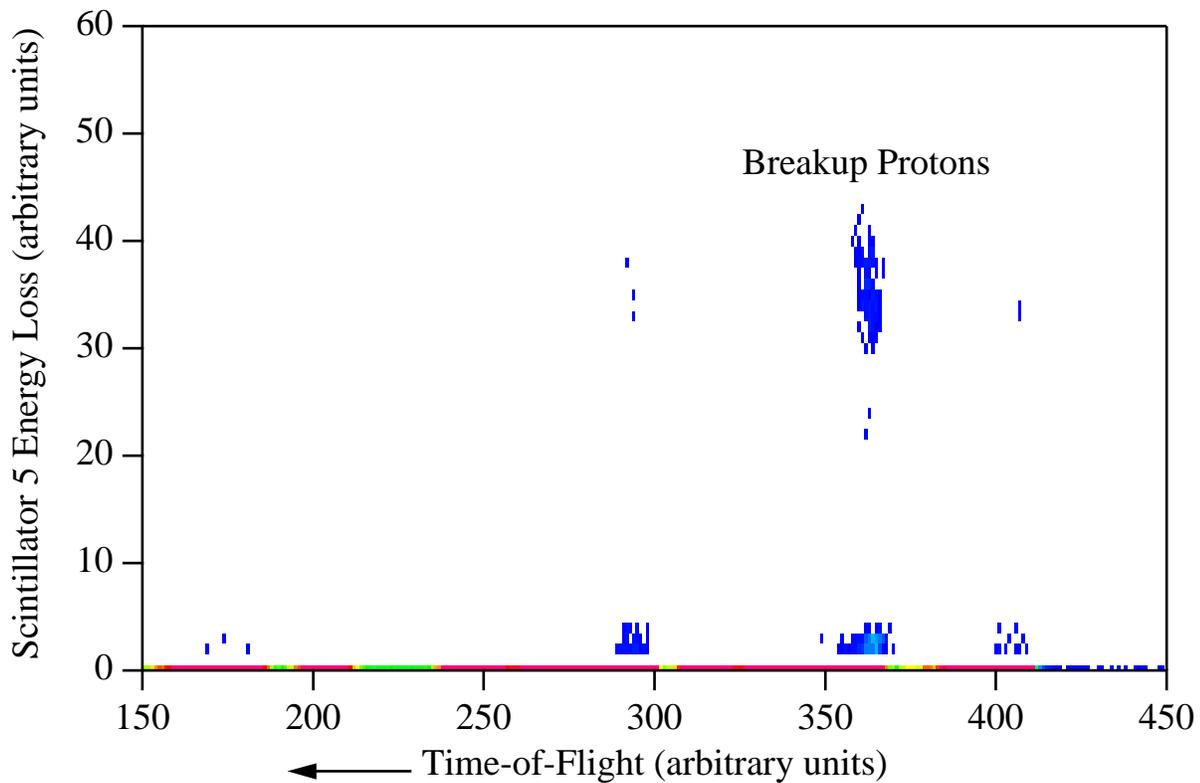} \caption{Typical proton scintillator bar
energy loss versus time-of-flight spectrum. The events with small scintillator
signals represent crosstalk from adjacent scintillator bars.} \label{ppid}
\end{figure}

\begin{figure}\epsfig{file=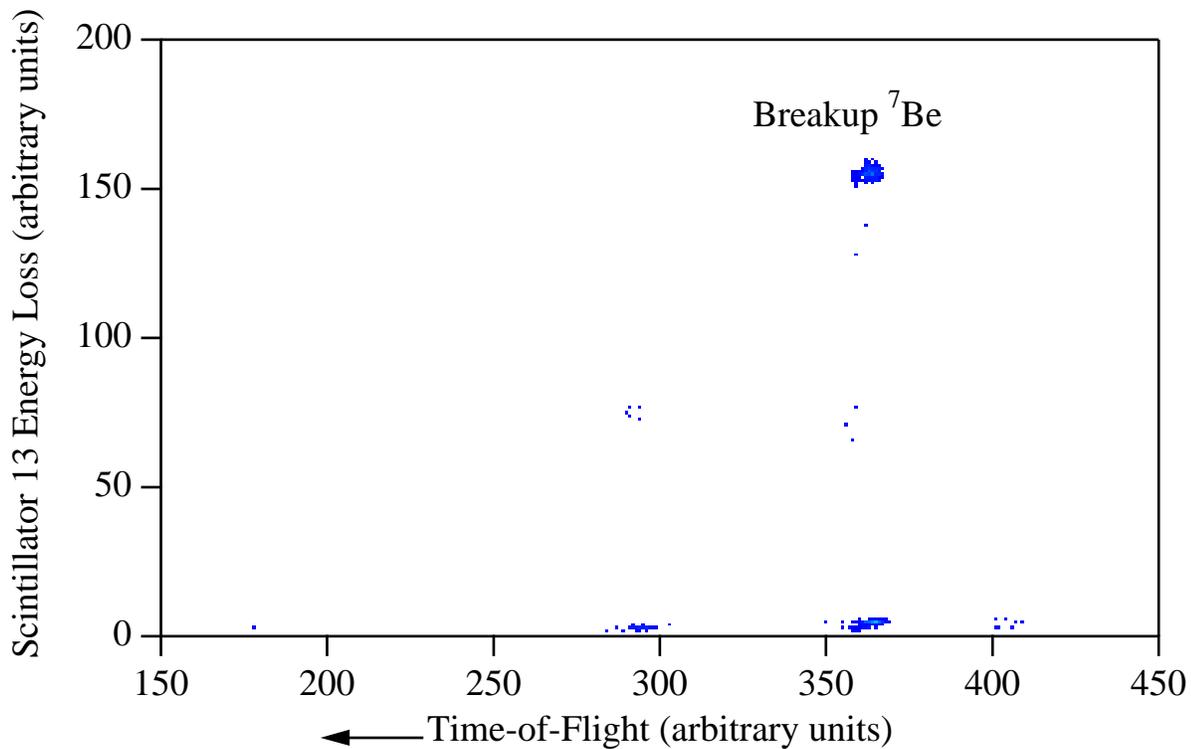} \caption{Typical $^{7}$Be scintillator bar
energy loss versus time-of-flight spectrum. The events with small scintillator
signals are due to light produced in adjacent scintillator bars.} \label{bepid}
\end{figure}

Particle identification was achieved through measurements of energy loss in the
plastic scintillator array and time-of-flight between the exit of the A1200 and
the scintillator array. The geometric average of signals from photomultiplier
tubes on the top and bottom of each scintillator bar served as a measure of
particle energy loss. Since the scintillator array was not sufficiently thick to
stop the breakup fragments, direct total energy measurements were not possible,
and the time-of-flight measurement was crucial for ion identification.
Calibration beams of $^{7}$Be and $\textit{p}$ having the same magnetic rigidity
as the $^{8}$B beam were used to confirm the particle identification. The
protons and $^7$Be struck widely separated scintillator bars, allowing
optimization of the individual bar electronics for the appropriate fragment
energy losses. Fig.\ \ref{ppid} shows the scintillator energy loss versus
time-of-flight spectrum for a scintillator bar that detected protons, while
Fig.\ \ref{bepid} shows that for a bar used to detect $^{7}$Be fragments. These
spectra are gated, requiring a good position signal in at least one proton MWDC
plane and one $^{7}$Be MWDC plane.

We reconstructed the 4-momenta of the breakup fragments from the measured
positions in all 8 MWDC planes and the magnetic field using the ion optics code
COSY INFINITY \cite{cosy2}. The magnetic field was measured with a Hall probe at
2184 points in each of 4 planes in the gap of the dipole magnet to a precision
of $\pm$ 2 mG \cite{jon}. Second-order Taylor series expansions about a
reference trajectory were employed, and the trajectory reconstruction was
checked through the use of proton and $^7$Be calibration beams. The invariant
mass method was used to calculate the relative energy of the fragments according
to $E_{rel} = \sqrt{E^{2}-\mathbf{p}^{2}}-m_{Be}c^{2}-m_{\textit{p}}c^{2}$,
where $E$ is the total relativistic energy, and $\mathbf{p}$ the total momentum
in the laboratory frame. The energy and momentum are defined by $E = \gamma_{Be}
m_{Be}c^{2} + \gamma_{\textit{p}} m_{\textit{p}}c^{2}$, and $\mathbf{p} =
\it{\gamma_{Be} m_{Be}} \mathbf{v}\it{_{Be}} + \it{\gamma_{\textit{p}}
m_{\textit{p}}} \mathbf{v_{\textit{p}}}$. The relative energy is simply the
kinetic energy in the center-of-mass reference frame. We obtained a relative
energy resolution of 55 keV (1$\sigma$) at $E_{rel}$ = 100 keV; the energy
resolution increased for higher relative energies approximately as
$\sqrt{E_{rel}}$. The small separation between the 1st and 2nd MWDCs, combined
with the MWDC position resolution, caused the angular resolution to limit the
relative energy resolution. This small distance was necessitated by the
requirement that the first detector be far enough away from the magnet that
there be adequate separation between the $^{7}$Be fragments and the $^{8}$B
beam, and by the limitations of an existing vacuum chamber. Other contributions
to the relative energy resolution included energy loss and multiple scattering
in the 47 mg cm$^{-2}$ target and the MWDCs, each of which was 30 mg cm$^{-2}$
thick.

The resolution and efficiency of the experimental apparatus were determined
through a Monte Carlo simulation. The inputs to the simulation included the beam
emittance (6 mm beam spot diameter, $\pm$ 6 mrad in the dispersive direction of
the magnet, $\pm$ 9 mrad in the non-dispersive direction) and the measured
detector position resolution. The beam emittance was measured by reducing the
magnetic field, causing the $^8$B beam to miss the beam blocker and be detected
in the MWDCs, while the detector position resolution was determined through the
use of a mask. The Monte Carlo simulation was also used to calculate the small
fraction of the $^7$Be breakup fragments that were intercepted by the beam
blocker.

In order to evaluate the geometric efficiency of the setup, we employed a model
for the breakup of $^8$B that includes both $E$1 and $E$2 transition amplitudes,
which have different distributions in $\Theta_{^{8}B}$, the laboratory
scattering angle of the excited $^8$B. To account for the $E$1-$E$2 interference
observed in the asymmetry of the longitudinal momentum distribution of $^7$Be
fragments, we included an anisotropic angular distribution of the breakup
fragments in the excited $^8$B rest frame. The shape of this distribution is
similar to those shown in Fig. 9 of ref.\ \cite{esbensen}, but was empirically
adjusted to reproduce the longitudinal momentum distribution of protons measured
in this experiment, which will be discussed in section\ \ref{results}. The $E$1
and $E$2 dissociation probabilities were taken from the model of ref.\
\cite{esbensen}, after scaling the $E$2 matrix elements by the factor 0.7. This
quenching of the $E2$ amplitudes, required for the best fit of the inclusive
data, is discussed in more detail below. We gauged the model-dependence of the
efficiency determination by also computing the efficiency using the same model
without $E$2 transitions. The difference between the computed efficiencies with
and without $E$2 transitions was less than 5\% for the angular and relative
energy ranges covered in the experiment. This difference was used as the
theoretical uncertainty in the efficiency determination.

\begin{figure}\epsfig{file=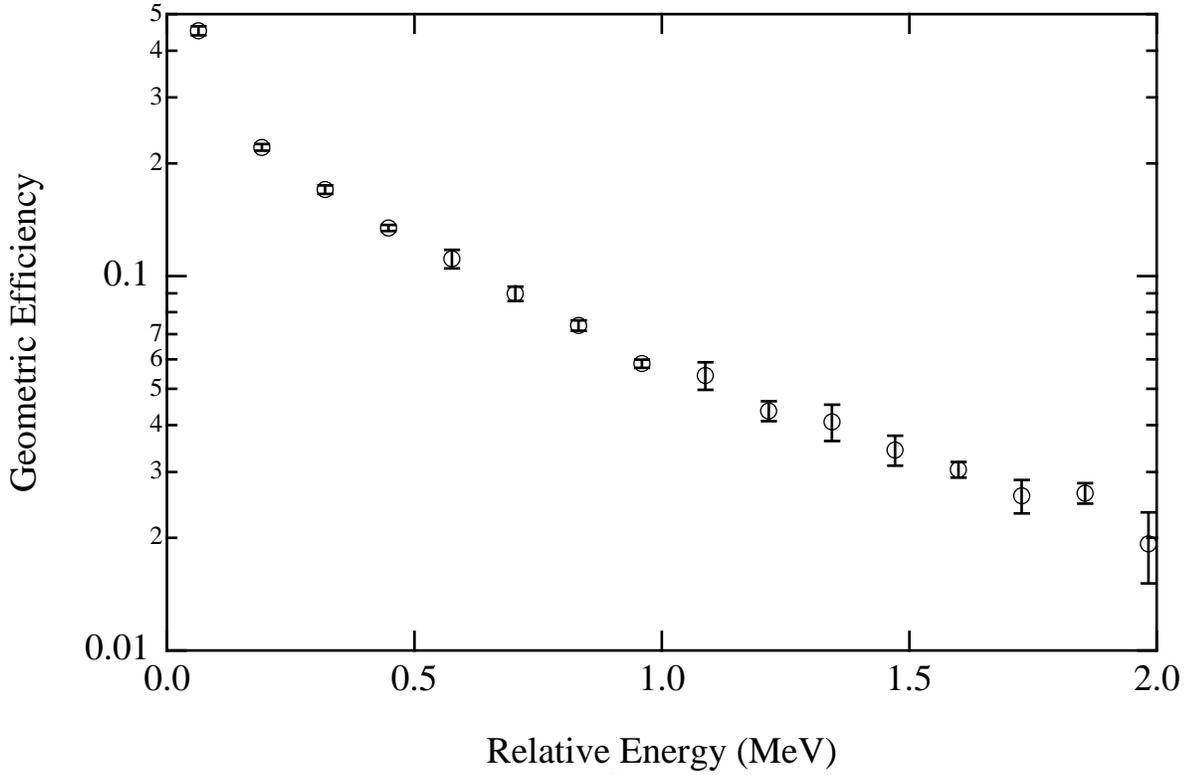} \caption{Geometric efficiency for detecting
protons and $^7$Be fragments in coincidence from the Coulomb dissociation of 83
MeV/nucleon $^8$B with impact parameters $\geq$ 30 fm. The relative errors shown
are statistical uncertainties from the simulation and theoretical uncertainties
from the size of the $E$2 component, added in quadrature.} \label{eff}
\end{figure}

Since both $E$2 transitions and nuclear absorption and diffraction effects are
relatively more important at small impact parameters than at large ones, we
imposed a impact parameter cutoff at 30 fm. For 83 MeV/nucleon $^8$B on Pb, this
corresponds classically to $\Theta_{^{8}B}$ = 1.77$^\circ$. In practice,
$\Theta_{^{8}B}$ was determined from the reconstructed total laboratory momentum
vector, and the 1$\sigma$ resolution of this quantity was 4.5 mrad. The
geometric efficiency for detecting $^8$B breakups with b $\geq$ 30 fm is shown
in Fig.\ \ref{eff}. The efficiency falls off rapidly with increasing relative
energy, primarily due to the small solid angle subtended by the proton MWDCs. As
the goal of the experiment was to determine the Coulomb dissociation cross
section at low relative energies, the experimental arrangement was most
sensitive to the events of interest. The intrinsic detection efficiency, i.e.,
the probability that all 8 MWDC planes provided good position signals when the
breakup fragments passed through them, was measured to be 0.414 $\pm$ 0.008
using the scintillator array.

\section{Results and Analysis}

\subsection{Theoretical Methods}

We have performed both 1st-order perturbation theory and continuum-discretized
coupled channels (CDCC) calculations of the Coulomb breakup of $^8$B. The
perturbation theory calculations have been described previously
\cite{esbensen}; we include a brief description of the CDCC calculations. Both
of these calculations assume a simple, single-particle potential model for the
structure of $^8$B: a $p_{3/2}$ proton coupled to an inert 3/2$^-$ $^7$Be
core. In the CDCC approach \cite{cdcc}, the breakup of $^8$B is assumed
to populate a selected set of spin-parity excitations with proton-$^{7}$Be
relative energies up to some maximum value. This excitation energy range is
subdivided into a number of intervals, or bins. For each such bin a
representative square integrable wave function is constructed, a superposition
of those proton-$^{7}$Be scattering states internal to the bin. These bin wave
functions form an orthonormal basis for the expansion and coupled channels
solution of the proton + $^{7}$Be + target three-body wave function. The $^8$B
and $^8$B$^*$ coupling potentials with the target are constructed by numerically
folding the proton-$^7$Be relative motion states with $U(\vec{r}, \vec{R})$, the
sum of the assumed interactions of the proton and $^7$Be with the target, which
is expanded to a maximum specified multipole order $\lambda$. This coupled
channels solution is carried out here using the code {\sc fresco} \cite{fresco}.
The subsequent evaluation of the fragment energy and angular distributions, from
the CDCC bin-state inelastic amplitudes, is discussed in detail in ref.\
\cite{jat00}.

The parameter space used in the CDCC calculations is as follows. Partial waves
up to $L_{max}=15000$ and radii up to 1000 fm were used for the computation of
the projectile-target relative motion wave functions.  The wave functions for
each bin and their coupling potentials were calculated using proton-$^{7}$Be
separations up to 200 fm. Excitations up to a proton-$^{7}$Be relative energy of
10 MeV were considered. In these calculations the $^7$Be intrinsic spin is
neglected, assuming that the core behaves as a spectator. The proton spin
dependence is included, however, and all proton-$^{7}$Be relative motion
excitations consistent with orbital angular momenta $\ell \leq 3$, i.e.\
relative motion states $\ell_j$ up to $f_{7/2}$, were included. The effects of
the $g$-wave continuum are small and are neglected. The calculations use
potential multipoles $\lambda \leq 2$ in the expansion of the proton- and
$^7$Be- target interactions.  The real potential used to construct the wave
functions for each bin was the same as that used to bind the $^8$B ground state,
a pure $p_{3/2}$ proton single-particle state. This proton-$^7$Be binding
potential was taken from Esbensen and Bertsch \cite{esbensen}, and was used for
all spin-parity channels. The fragment-target nuclear interactions are also
included; for the $^7$Be-$^{208}$Pb system we take the ($^7$Li) interaction of
Cook \cite{cook} and for the proton-$^{208}$Pb system the global nucleon optical
potential of Becchetti and Greenlees \cite{becc}.

\subsection{Longitudinal Momentum Distributions} \label{results}

\begin{figure}\epsfig{file=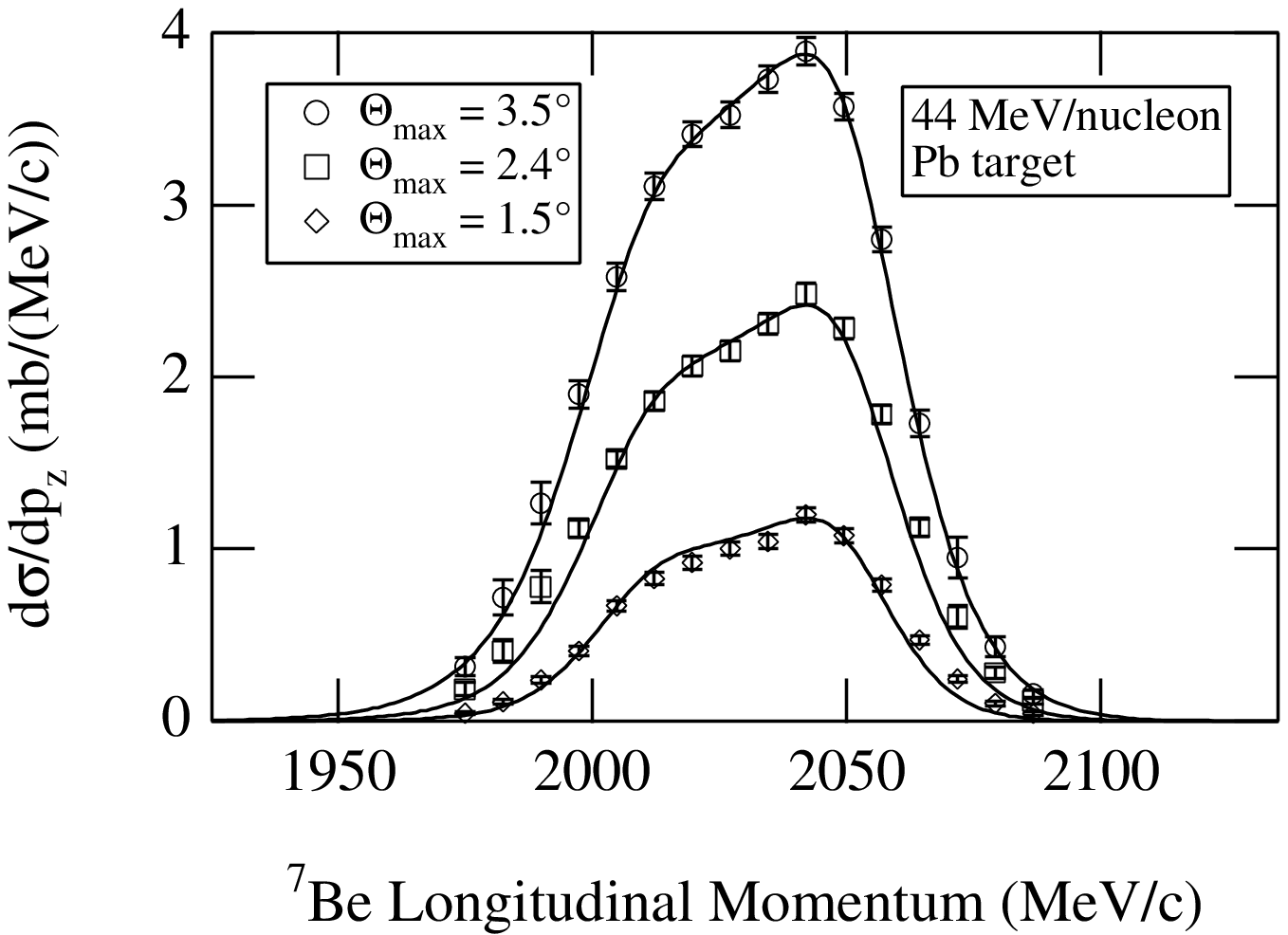} \caption{Measured longitudinal momentum
distributions of $^7$Be fragments from the Coulomb dissociation of 44
MeV/nucleon $^8$B on Pb with several maximum $^7$Be scattering angle cuts. Also
shown are 1st-order perturbation theory calculations convoluted with the
experimental resolution. See the text for details.} \label{45pblmd} \end{figure}

Measured laboratory frame longitudinal momentum distributions of $^7$Be
fragments from the Coulomb breakup of 44 MeV/nucleon $^8$B on a Pb target are
shown in Fig.\ \ref{45pblmd}. The momentum resolution obtained was 5 MeV/c, and
the error bars indicate the relative uncertainties of the data points, which are
dominated by statistical errors. The systematic uncertainty in the measured
cross section due to target thickness and beam intensity was $\pm$ 9\%. This
systematic uncertainty is common to all of the $^7$Be momentum distribution
measurements. Fig.\ \ref{45pblmd} also includes the results of 1st-order
perturbation theory calculations performed using a modified version of the model
of ref.\ \cite{esbensen}. Both the overall normalization and the $E2$ matrix
elements of this calculation have been scaled, the former by 1.22 and the latter
by 0.7. We shall return to this point later.

\begin{figure}\epsfig{file=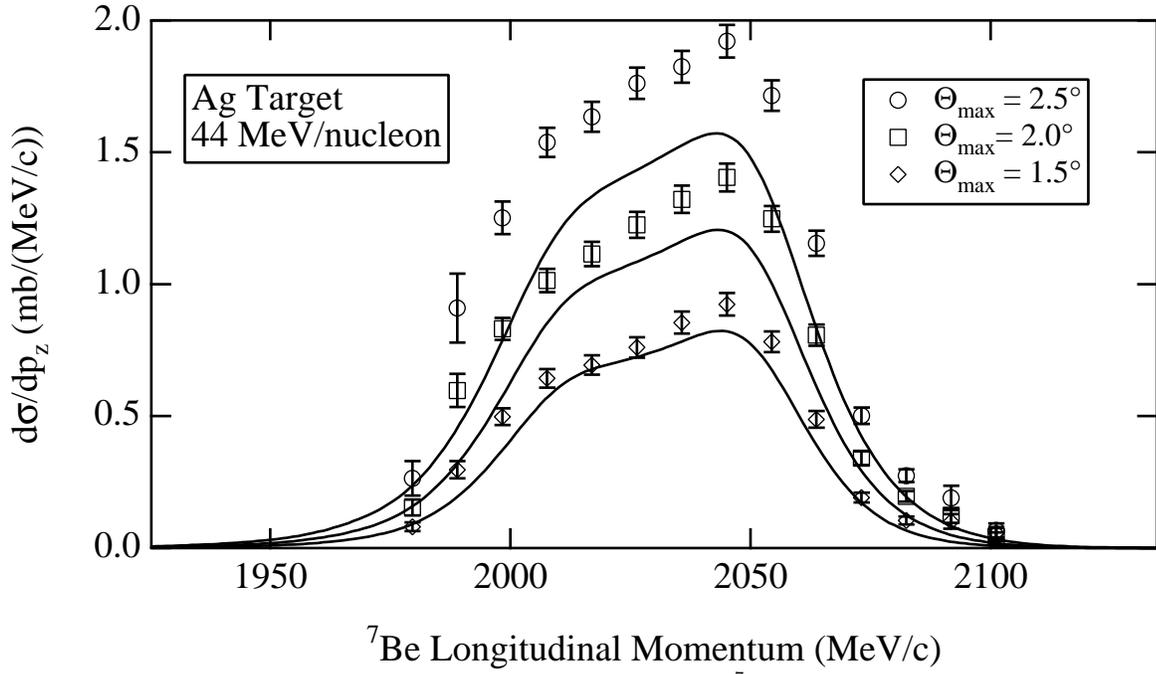} \caption{Laboratory frame longitudinal
momentum distributions of $^7$Be fragments with maximum scattering angles of
2.5$^\circ$, 2.0$^\circ$, and 1.5$^\circ$ emitted in the breakup of 44
MeV/nucleon $^8$B on Ag. The solid curves represent perturbative Coulomb
dissociation calculations convoluted with the experimental resolution.}
\label{45aglmd} \end{figure}

To investigate any possible dependence of higher-order electromagnetic effects
on target charge, we also made inclusive measurements with an Ag target. The
measured longitudinal momentum distributions of $^7$Be fragments produced in the
dissociation of 44 MeV/nucleon $^8$B on Ag are shown in Fig.\ \ref{45aglmd} for
several different maximum $^7$Be scattering angle cuts. The agreement with the
1st-order perturbation theory calculations done with the model of ref.\
\cite{esbensen} shown here is not as good as with the Pb target. In particular,
the magnitude and width of the calculations are insufficient to describe the
data. These 1st-order perturbation theory calculations have the same $E2$ matrix
element scaling and overall normalization as the 44 MeV/nucleon Pb target
calculations. It is possible that nuclear processes not accounted for in the
Coulomb dissociation calculation are responsible for this discord. The
measurement of ref.\ \cite{kelley} of nuclear-induced breakup of 41 MeV/nucleon
$^8$B on a Be target found a symmetric longitudinal momentum distribution. The
difference between the Coulomb dissociation calculations and the data increased
with maximum scattering angle, consistent with an increasing relative importance
of nuclear-induced breakup. The breakup of $^8$B on Ag can be studied with the
CDCC method, but these results are outside the scope of this paper, and will be
presented elsewhere.

\begin{figure}\epsfig{file=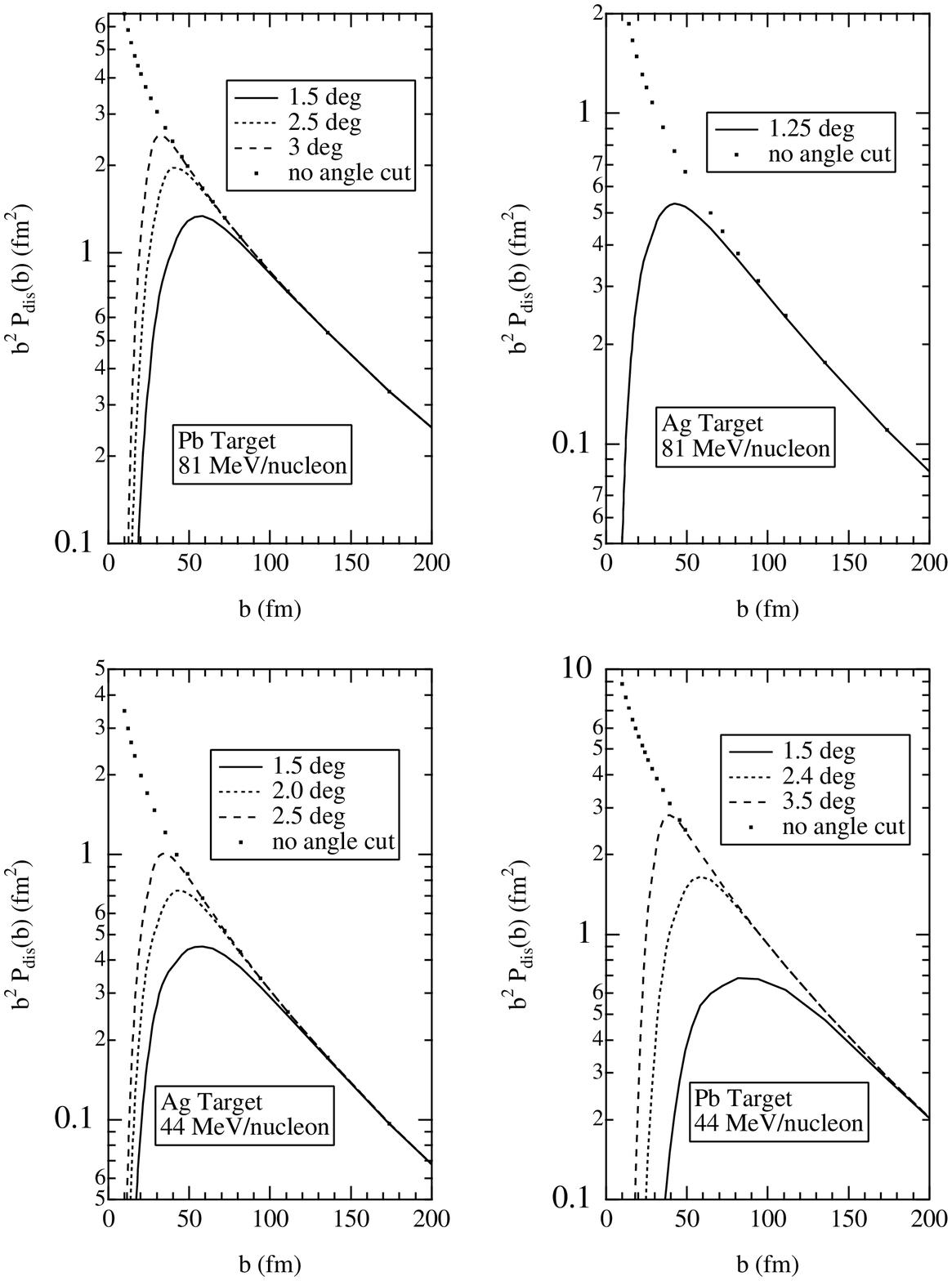,width=7in} \caption{Product of Coulomb
dissociation probability and impact parameter squared for the breakup of $^8$B
with various $^7$Be scattering angle cuts as a function of impact parameter.
This is a measure of the relative $^7$Be detection probability, revealing the
impact parameter sensitivity of the various angle cuts.} \label{bsens}
\end{figure}

Placing different cuts on the angles of the emitted $^7$Be fragments allows one
to probe different impact parameters. However, a maximum $^7$Be scattering angle
does not correspond to a fixed minimum impact parameter, because the breakup
energy and the angle of the emitted proton are not determined in the inclusive
measurement. The sensitivities of the various angular cuts of the longitudinal
momentum distributions to different impact parameters are shown in Fig.\
\ref{bsens}. These curves are a measure of the relative probability that $^7$Be
fragments emitted in Coulomb breakups at various impact parameters will fall
within specified angular cuts. All of these calculations were performed using
the model of ref.\ \cite{esbensen}, with the $E2$ matrix elements scaled by 0.7.
A comparison of the figures reveals that the Ag distributions probe smaller
impact parameters than the Pb distributions, indicating that nuclear absorption
and diffraction should play a larger role for the Ag target.

\begin{figure}\epsfig{file=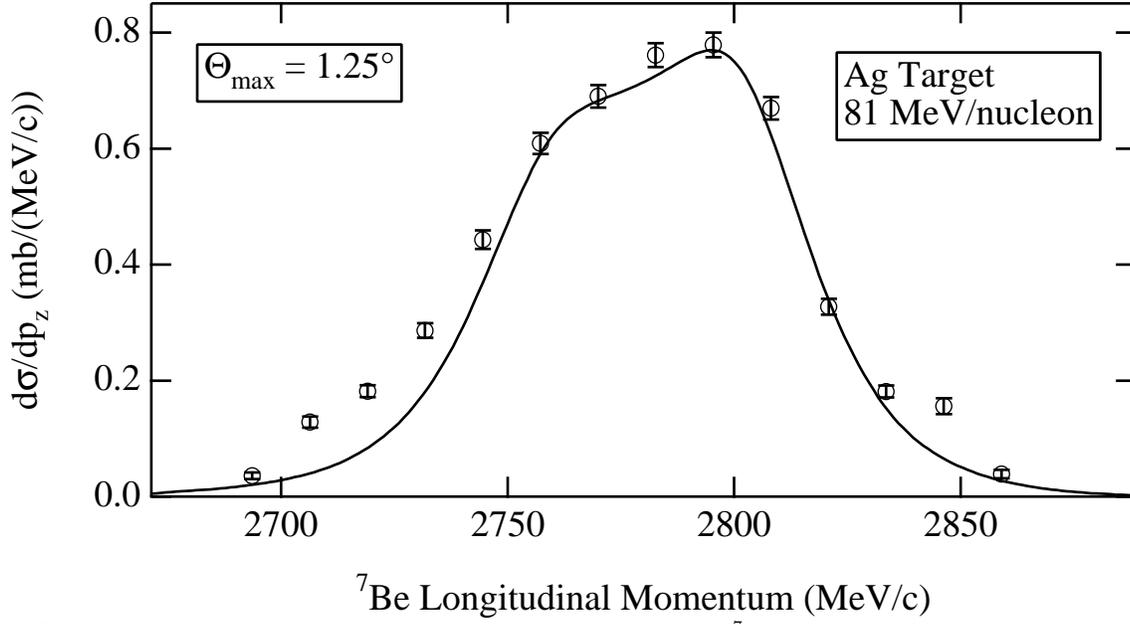} \caption{Laboratory frame longitudinal
momentum distribution of $^7$Be fragments with maximum scattering angles of
1.25$^\circ$ emitted in the Coulomb dissociation of 81 MeV/nucleon $^8$B on Ag.
The curve is a 1st-order perturbation theory calculation convoluted with the
experimental resolution.} \label{81aglmd} \end{figure}

Fig.\ \ref{81aglmd} shows the $^7$Be longitudinal momentum distribution for the
81 MeV/nucleon $^8$B beam on Ag with a maximum $^7$Be scattering angle cut of
1.25$^\circ$. The curve is a 1st-order perturbation theory calculation done with
the model of ref.\ \cite{esbensen} with $E2$ matrix elements scaled by 0.7. The
overall normalization of this calculation has not been altered. The perturbative
calculation describes the data fairly well, with the most important discrepancy
being the greater width of the measured distribution. It is possible that
nuclear absorption and diffraction not accounted for in the Coulomb dissociation
calculation broaden the measured distribution beyond the predicted extent (see
Fig.\ \ref{bsens}).

\begin{figure}\epsfig{file=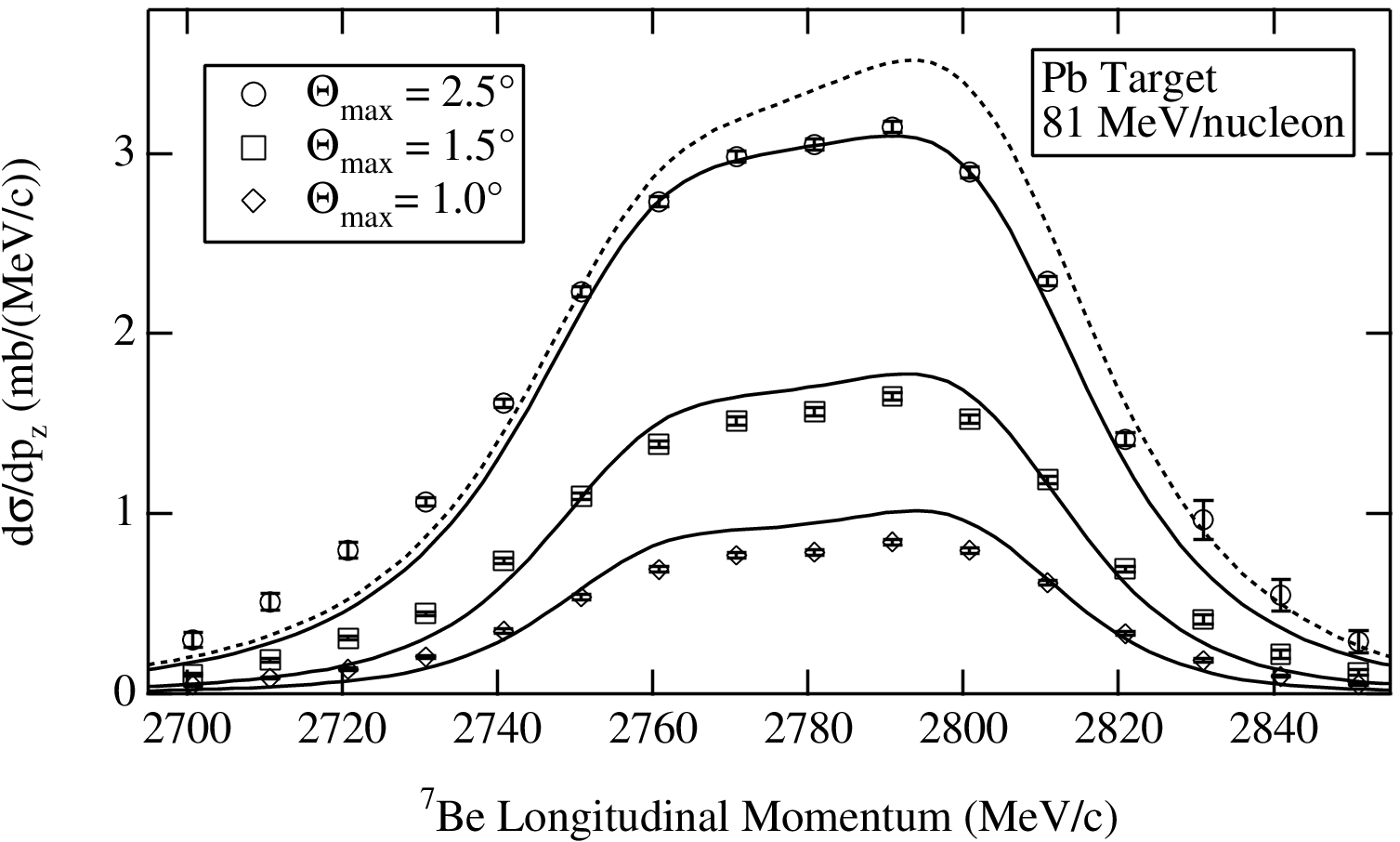} \caption{Laboratory frame longitudinal
momentum distribution of $^7$Be fragments emitted in the Coulomb dissociation
of 81 MeV/nucleon $^8$B on Pb with maximum scattering angles of 2.5$^\circ$,
1.5$^\circ$, and 1.0$^\circ$. The solid curves are continuum-discretized
coupled channels calculations that include both Coulomb and nuclear
interactions, convoluted with the experimental resolution. The dashed curve is
a DWBA calculation for $\Theta_{max}$ = 2.5$^\circ$.} \label{81pblmd}
\end{figure}

The inclusive $^7$Be longitudinal momentum distributions measured at 81
MeV/nucleon with the Pb target are depicted in Fig.\ \ref{81pblmd} for $^7$Be
scattering angle cuts of 2.5$^\circ$, 1.5$^\circ$, and 1.0$^\circ$. Also shown
here are CDCC calculations convoluted with the experimental resolution of 5
MeV/c. The CDCC calculations describe the data reasonably well, accurately
reproducing the slopes of the central regions of the momentum distributions,
particularly for the largest angle cut. These calculations are not fits, but
rather are absolute predictions based on the assumed structure model; the $E1$
and $E2$ matrix elements have not been scaled in the CDCC calculations. The
dashed curve is a distorted wave Born approximation (DWBA) calculation for the
largest angle cut that assumes the same structure model and interactions as the
CDCC calculation. The difference between the 1st-order DWBA and the CDCC
calculations reflects the influence of higher-order processes, which tend to
reduce the effective $E2$ strength needed in the 1st-order calculation. As is
the case for the 81 MeV/nucleon Ag data, the calculations predict distributions
narrower than were measured. The difference in magnitude between the
calculations and the data for the smaller angle cuts is within the error due to
the angular uncertainty of 0.25$^\circ$. This angular uncertainty is common to
all the angle cuts, but has a greater influence on the uncertainty in the
magnitude of the cross section for the smaller angle cuts, as it represents a
larger fraction of the total angular coverage.

\begin{figure}\epsfig{file=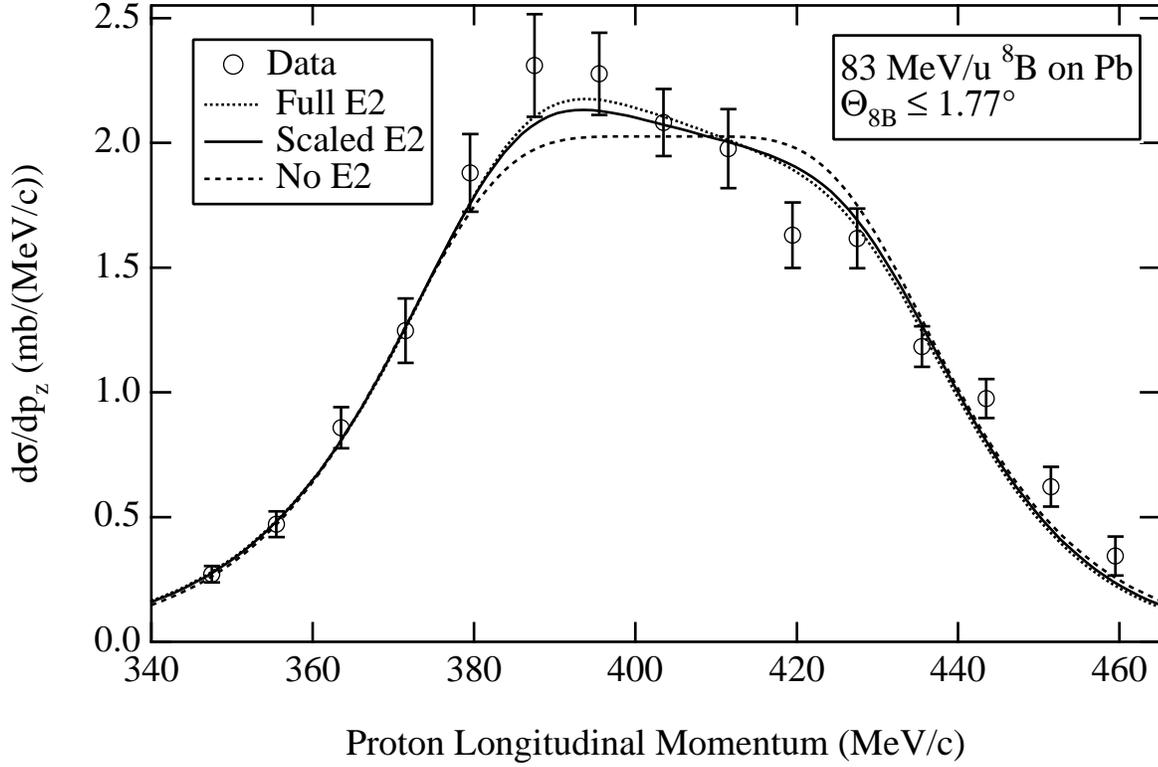} \caption{Measured longitudinal momentum
distribution of protons from the Coulomb dissociation of 83 MeV/nucleon $^{8}$B
on Pb with $^{8}$B scattering angles $\leq 1.77^{\circ}$. Only relative errors
are shown. Also depicted are 1st-order perturbation theory calculations with
different E2 strengths, convoluted with the experimental resolution.}
\label{plmd} \end{figure}

Fig.\ \ref{plmd} depicts the longitudinal momentum distribution of protons
measured in coincidence with $^7$Be fragments from the Coulomb breakup of 83
MeV/nucleon $^8$B with reconstructed $^8$B center-of-mass angles of 1.77$^\circ$
and less. The proton momentum resolution was estimated from the Monte Carlo
simulation to be 4~MeV/c (1$\sigma$). Also shown in the figure are 1st-order
perturbation theory calculations using the model of ref.\ \cite{esbensen}, one
with the full $E2$ amplitude, one with the $E2$ matrix elements scaled by 0.7,
and another with no $E2$ matrix elements. The three calculations were
renormalized by 10\% or less in order to facilitate comparison.

All of the measured longitudinal momentum distributions share a common feature:
an asymmetry attributed to interference between $E$1 and $E$2 transition
amplitudes in the Coulomb breakup. This effect was first predicted for
$^8$B breakup in ref.\ \cite{bertsch}, though its importance in the Coulomb
breakup of Li and O projectiles was noted earlier \cite{baur89}. An early
measurement of the momentum distribution of $^7$Be fragments from the
Coulomb breakup of 41 MeV/nucleon $^8$B on gold \cite{kelley} provided
evidence of this effect, but the statistics were insufficient to draw any
definitive conclusions. By measuring longitudinal momentum distributions of
$^7$Be nuclei and protons on two targets at two different beam energies
with two different experimental setups, we have conclusively demonstrated
the existence of this asymmetry.

\begin{figure}\epsfig{file=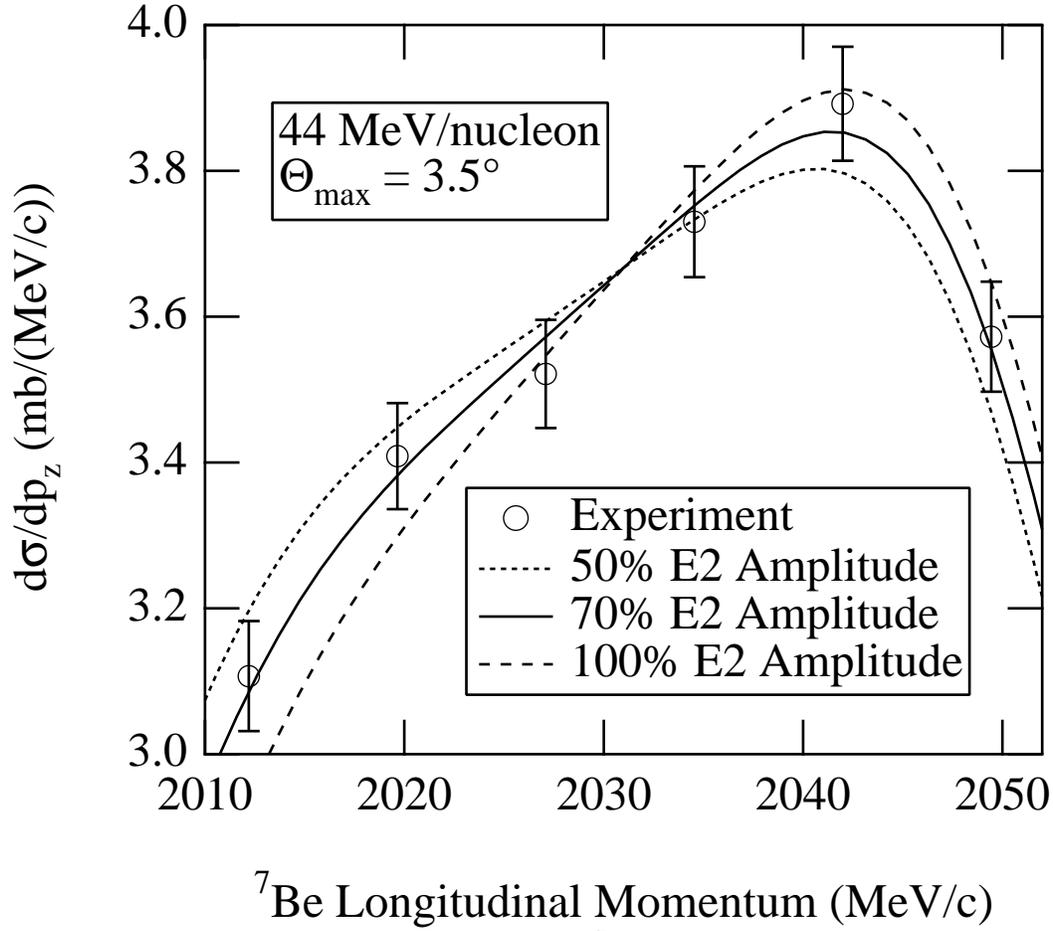} \caption{Central region of the
3.5$^{\circ}$ angle cut of the $^7$Be longitudinal momentum distribution from
the breakup of 44 MeV/nucleon $^8$B on Pb. The curves are calculations performed
with different E2 matrix elements, expressed in terms of the $E2$ amplitude of
the model of \protect\cite{esbensen}, normalized to the center of the
distribution.} \label{e2comp} \end{figure}

In 1st-order perturbation theory, the size of the predicted asymmetry is
proportional to the $E2$ transition amplitude.  Fig.\ \ref{e2comp} illustrates
this point, depicting the central region of the 3.5$^\circ$ 44 MeV/nucleon
$^7$Be longitudinal momentum distribution from the Pb target along with three
calculations. These calculations were performed with different $E2$ amplitudes,
and are normalized to the same value at the center of the distribution. The
simple potential model of $^8$B structure from ref.\ \cite{esbensen} makes
predictions for the $E1$ and $E2$ matrix elements. By arbitrarily scaling the
$E1$ and $E2$ matrix elements in a 1st-order perturbation theory of the Coulomb
breakup, we fit the central 6 data points of the 3.5$^\circ$ $^7$Be longitudinal
momentum distribution from the breakup of 44 MeV/nucleon $^8$B on the Pb target
in order to minimize the $\chi^2$ value. Using this procedure, we found that the
optimal ratio of $E2$ and $E1$ matrix element scaling factors was 0.7. The same
ratio of matrix element scaling factors was required to best fit the 81
MeV/nucleon $^7$Be longitudinal momentum distribution on Pb in perturbation
theory, a calculation that is not shown here (see ref.\ \cite{davids}). In the
exclusive experiment, it was not possible to measure the longitudinal momentum
distributions with a precision comparable to that of the inclusive measurement.
Furthermore, any nuclear-induced breakup contribution is relatively more
important for the Ag target than for the Pb target. For these reasons, we used
only the inclusive measurements on Pb to deduce the $E2$ strength. The
preliminary findings of the inclusive measurement were described previously
\cite{davids}. In ref.\ \cite{davids}, the optimal ratio of the $E2$ and $E1$
matrix element scaling factors was incorrectly reported as the ratio of the
scaling factors for the $E2$ and $E1$ strength distributions; the correct value
for this ratio is 0.7$^{2}$ = 0.49.  As a consequence, the reported
\cite{davids} ratio of $E2$ and $E1$ S factors at $E_{rel}$ = 0.6 MeV should be
replaced by 4.7 $^{+ 2.0}_{- 1.3} \times 10^{-4}$. This result assumes the
validity of 1st-order perturbation theory in describing the reaction mechanism,
and a particular $^8$B structure model. If higher-order electromagnetic effects
are important, a larger intrinsic $E2$ strength is required to fit the data.
Hence we have determined the effective $E2$ matrix element which, within a
1st-order perturbation theory with a given $E1$ matrix element, fits the
empirically observed asymmetry in the longitudinal momentum distributions.

Table\ \ref{table2} lists the integrated cross sections obtained in the
inclusive longitudinal momentum distribution measurements on both targets. The
purpose of these inclusive measurements was to deduce the $E2$ strength in the
Coulomb breakup. A determination of low-lying $E1$ strength would be subject to
large nuclear structure uncertainties, since the inclusive measurements are
sensitive to electromagnetic strength over a large range of excitation energies.
However, the observed asymmetry in the longitudinal momentum distributions is a
clear signature of $E1$-$E2$ interference, through which these measurements
probe the total $E2$ strength.

\subsection{Breakup Energy Spectrum}

In contrast to the $E2$ component, the size of the $M1$ contribution to the
cross section for Coulomb breakup can be determined from the measurement of the
radiative capture cross section at the 0.64 MeV 1$^+$ resonance
\cite{filippone}. $M1$ transitions only play a role in Coulomb dissociation near
this energy, and the magnitude of the contribution is obtained from the measured
resonance parameters \cite{ajzenberg} and the calculated virtual photon spectrum
\cite{bertulani88}. The energy resolution of our exclusive measurement is too
large and the contribution too small to allow us to clearly see this resonance,
but it represents a few percent of the measured cross section.

Since the radiative capture reaction involves protons and $^7$Be nuclei in their
ground states, Coulomb breakup that yields excited $^7$Be nuclei is not relevant
to the inverse radiative capture rate. As our experimental setup did not include
any provision for detecting $\gamma$ rays, a correction for the yield to the
1/2$^-$ excited state of $^7$Be was made on the basis of equation 41 of ref.\
\cite{bertulani94}, and the analysis of the data of ref.\ \cite{kikuchi97} found
in ref.\ \cite{mengoni}. The size of this correction ranged from 1\% at 200 keV
to 9\% at 2 MeV.

\begin{figure}\epsfig{file=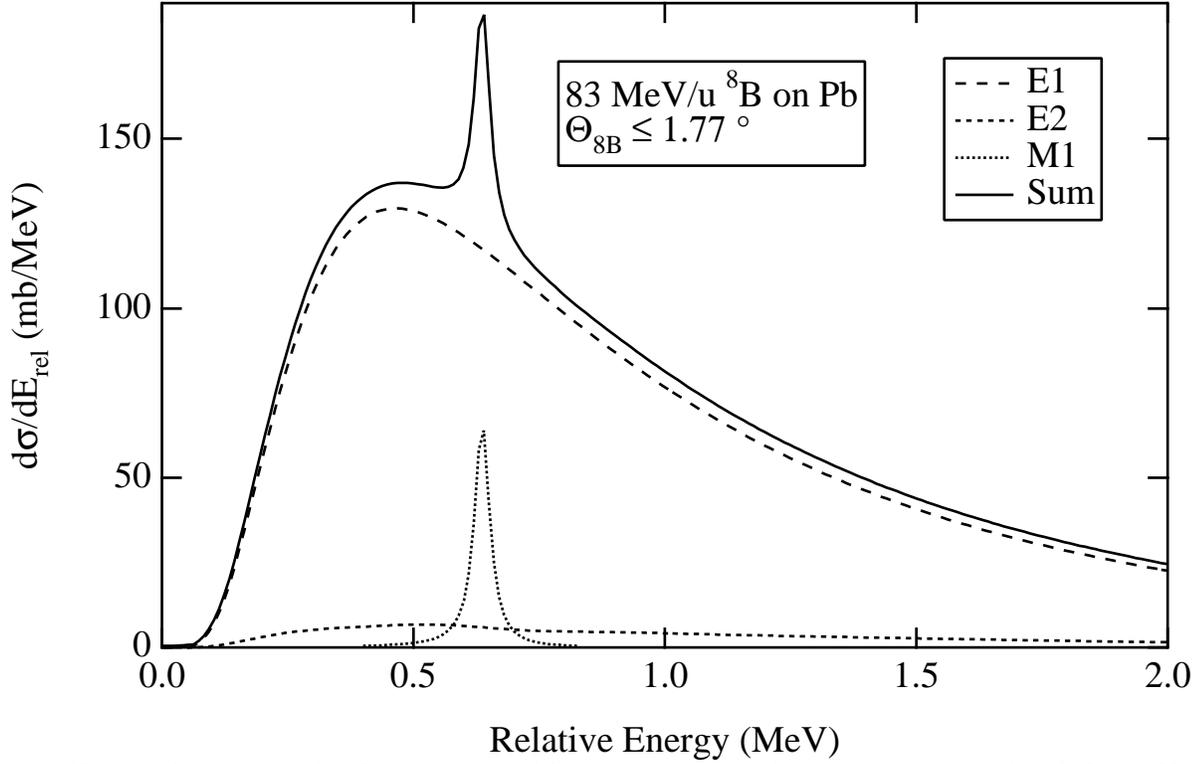} \caption{Contributions of $E1$, $E2$, and
$M1$ transitions to the cross section for the Coulomb dissociation of 83
MeV/nucleon $^{8}$B on Pb with $^{8}$B scattering angles $\leq 1.77^{\circ}$ in
1st-order perturbation theory. The $M1$ cross section is calculated by folding
the $M1$ $S$ factor measured in ref.\ \protect\cite{filippone} with the virtual
photon spectrum. The $E1$ and $E2$ cross sections are calculated using the model
of ref.\ \protect\cite{esbensen}, scaling the $E2$ matrix elements by the factor
0.7 required to reproduce the measured $^7$Be longitudinal momentum
distributions.} \label{thdes} \end{figure}

\begin{figure}\epsfig{file=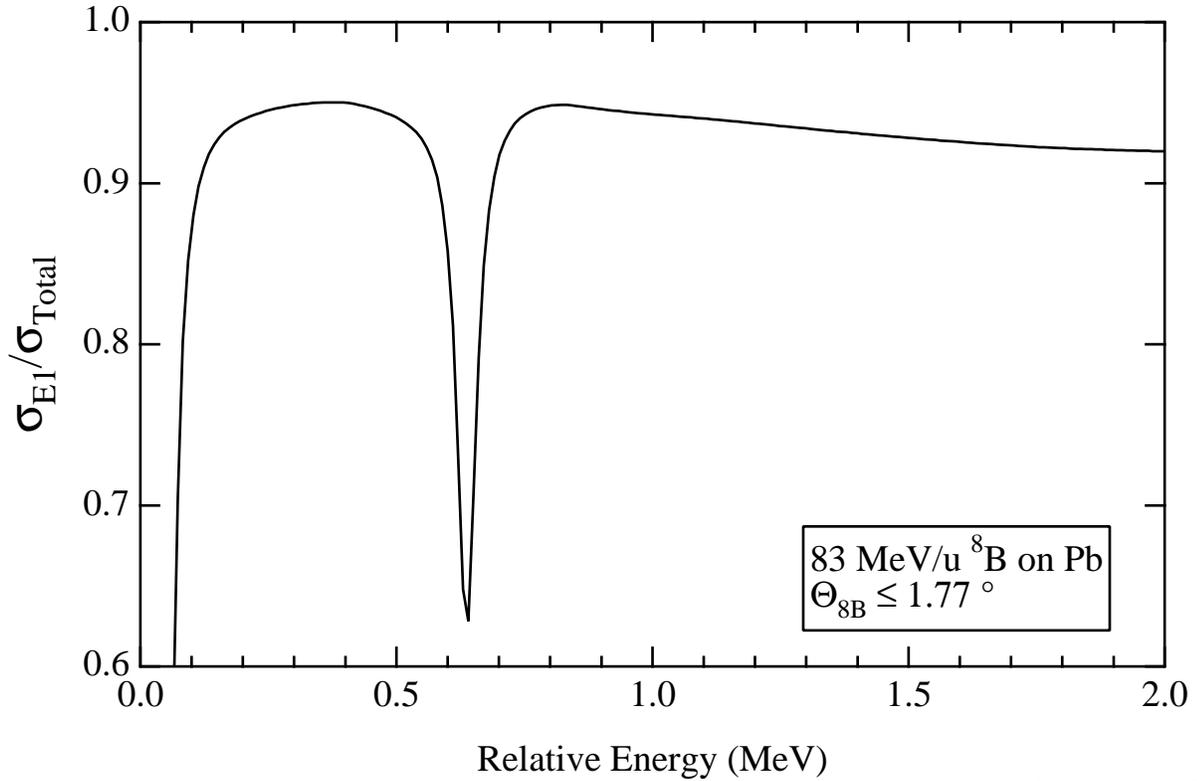} \caption{Fraction of the calculated cross
section for the Coulomb dissociation of 83 MeV/nucleon $^{8}$B on Pb with
$^{8}$B scattering angles $\leq 1.77^{\circ}$ (b~$\geq$~30~fm) accounted for by
$E1$ transitions in 1st-order perturbation theory. As the energy falls below 130
keV, $E2$ transitions become increasingly important.} \label{e1frac}
\end{figure}

Fig.\ \ref{thdes} shows the theoretical breakup energy spectrum calculated in
1st-order perturbation theory. The calculation was performed with the model of
ref.\ \cite{esbensen} for the $E1$ and $E2$ components, scaling the $E2$ matrix
elements by 0.7, while the $M1$ component was calculated as described above. By
placing a 1.77$^\circ$ cut on the reconstructed angle of the dissociated 83
MeV/nucleon $^8$B projectiles, we have ensured that nuclear diffraction and
absorption effects are small, and that the point-like projectile approximation
employed in 1st-order perturbation theory is valid. Furthermore, by also
excluding relative energies below 130 keV from our analysis, we minimized the
role of $E2$ transitions and considered only relative energies where $E1$ was
the dominant ($>$ 90\%) contribution to the breakup. As illustrated in Fig.\
\ref{e1frac}, $E1$ transitions dominate the breakup cross section from 130 keV
to 2 MeV except for a narrow range surrounding the 0.64 MeV $1^+$ resonance.
$E2$ transitions contribute significantly at relative energies under 130 keV,
accounting for the sharp fall in the $E1$ fraction of the cross section at low
relative energies.

To deduce the $E1$ strength at low relative energies, we carried out the
following procedure. After fixing the $E2$/$E1$ ratio using the inclusive data,
we convoluted the calculated $E1$, $E2$, and $M1$ cross sections with the
energy-dependent experimental resolution, and then scaled the combined $E1+E2$
cross section in order to minimize $\chi^2$ for the measured differential cross
section between 130 and 400 keV. Recent work \cite{jennings} suggests that above
400 keV, nuclear structure uncertainties increase appreciably. The best-fit
normalization factor of the $E1+E2$ calculation for the data between 130 keV and
400 keV was 0.93~$^{+ 0.05}_{- 0.04}$, resulting in $S_{17}(0)$~=~17.8~$^{+
1.4}_{- 1.2}$~eV~b, with all sources of uncertainty added in quadrature. We
extrapolated to zero energy using the prescription of Jennings {\em et al.}
\cite{jennings}. The quoted error (1$\sigma$) includes energy-dependent
contributions from statistics, momentum and angular acceptance, detector
efficiency, and the $^7$Be excited state yield correction, added in quadrature
with systematic uncertainties from the beam intensity (1\%), extrapolation to
zero energy (1\%), size of the $E2$ component (2.5\%), target thickness (2.6\%),
and momentum calibration accuracy (4.2\%).

We also analyzed the measured breakup cross section at higher relative energies,
carrying out the same $\chi^2$ minimization procedure for the data from 130 keV
to 2 MeV. The data above 2 MeV were excluded from the fit because of a 3$^{+}$
resonance at 2.2 MeV that was not included in the theoretical calculation, and
because the statistics there are poor. The best-fit normalization factor
obtained for the data between 130~keV and 2~MeV with this procedure was
1.00~$^{+ 0.02}_{- 0.06}$. We assign a 5\% theoretical extrapolation uncertainty
for this energy range \cite{jennings}. The result of the perturbation theory
analysis of data from 130~keV to 2~MeV is $S_{17}$(0)~=~19.1~$^{+ 1.5}_{-
1.8}$~eV~b. This result is consistent with the value extracted from the data up
to 400 keV, implying that the simple potential model of ref.\ \cite{esbensen}
describes the physics well even at large relative energies, within the
uncertainties. Nevertheless, we prefer the value of $S_{17}(0)$ inferred from
the data below 400 keV because of its relative insensitivity to the details of
$^8$B structure.

\begin{figure}\epsfig{file=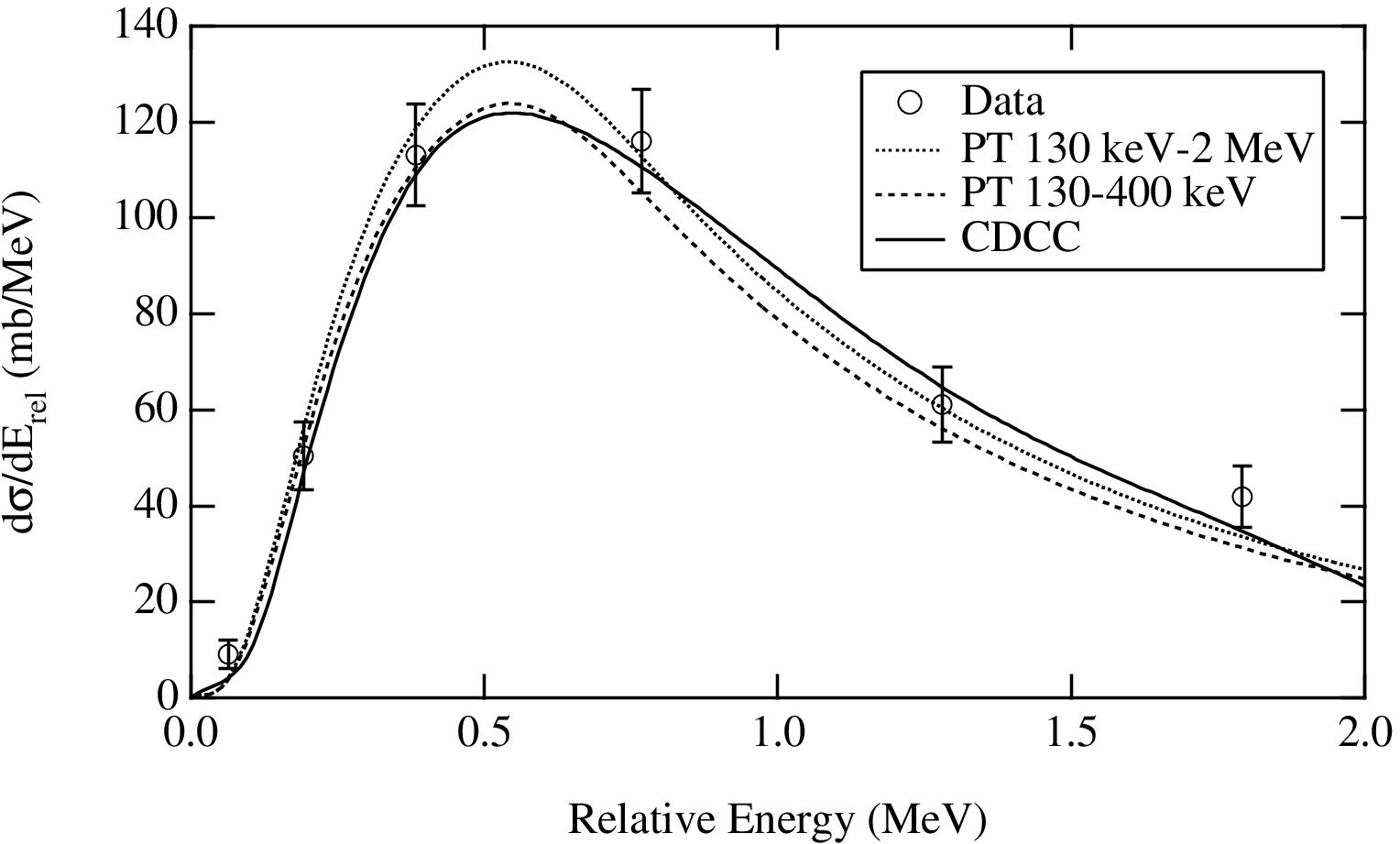} \caption{Measured differential cross section
for the Coulomb dissociation of 83 MeV/nucleon $^{8}$B on Pb with $^{8}$B
scattering angles $\leq 1.77^{\circ}$. Only relative errors are shown. Also
depicted are continuum-discretized coupled channels and two 1st-order
perturbation theory calculations, convoluted with the experimental resolution.
The point at 64 keV has been excluded from the fits because $E2$ transitions are
dominant at this energy.} \label{des} \end{figure}

Fig.\ \ref{des} shows the differential cross section measured in the exclusive
experiment along with the results of the best-fit 1st-order perturbation theory
calculations for the two energy ranges described above, performed using the
model of ref.\ \cite{esbensen} with $E2$ matrix elements quenched as required
to fit the inclusive data. The perturbation theory calculations include $M1$
transitions and have been convoluted with the experimental energy
resolution. The figure also includes the results of our CDCC calculations,
convoluted with the experimental resolution. The CDCC calculations employ a
slightly simplified version of the structure model of ref.\ \cite{esbensen},
and provide a means of gauging the importance of nuclear-induced breakup and
higher-order electromagnetic effects; the $E1$ and $E2$ reduced transition
probabilities predicted by the two structure models agree at the 1\% level.
These fully quantum mechanical CDCC calculations include both nuclear and
Coulomb interactions, and have not been renormalized.

The two reaction models describe the data between 130 keV and 2 MeV equally
well, implying that the theoretical uncertainties in the reaction mechanism are
smaller than or comparable to the experimental uncertainties here. In large
measure, this is due to the experimental conditions of the exclusive
measurement. By limiting the angular acceptance as we did, we probed large
impact parameters where the $E2$ and nuclear contributions are small. These CDCC
calculations indicate that nuclear-induced breakup is negligible at relative
energies less than 400 keV. Higher-order electromagnetic effects are also
smallest at the largest impact parameters \cite{baur96,esbensen}. The fact that
the zero energy $S$ factors implicit in the CDCC calculation (18.9 eV b) and the
best-fit 1st-order perturbation theory calculation for the data up to 2 MeV
(19.1 eV b) agree within 1\% gives confidence that 1st-order perturbation theory
adequately describes the underlying physics of the breakup reaction under these
experimental conditions, provided the $E2$ matrix elements are appropriately
quenched.

\section{Discussion}

The $E2$ strength deduced from the inclusive momentum distribution measurement
is 10 to 100 times larger than the upper limits reported in other experimental
studies \cite{iwasa,kikuchi97}. We studied an observable that directly probes
$E1$-$E2$ interference, the asymmetry in the longitudinal momentum distribution.
Our experimentally deduced value for the $E2$/$E1$ ratio is only slightly
smaller than or in good agreement with recent theoretical calculations
\cite{esbensen,typel,bennaceur,descouvement,barker}, and is consistent with the
measurement of \cite{guimaraes}, although this group does not give a value for
$S_{E2}/S_{E1}$. That the extracted experimental value should be somewhat
smaller than the theoretical values is consistent with the idea that 1st-order
perturbation theory overestimates the $E2$ contribution to the cross section
\cite{esbensen}. Table\ \ref{table3} shows the $E2$ strength predictions given
in several recent papers using potential models, microscopic cluster models, the
shell model embedded in the continuum, and R-matrix theory, along with the
results of this work. The concordance of these predictions of the $E2$ strength
made on the basis of disparate theoretical methods and the result deduced from
the measured longitudinal momentum distribution asymmetries imply that the $E2$
component must be accounted for in a proper theoretical description of the
Coulomb breakup.

\begin{figure}\epsfig{file=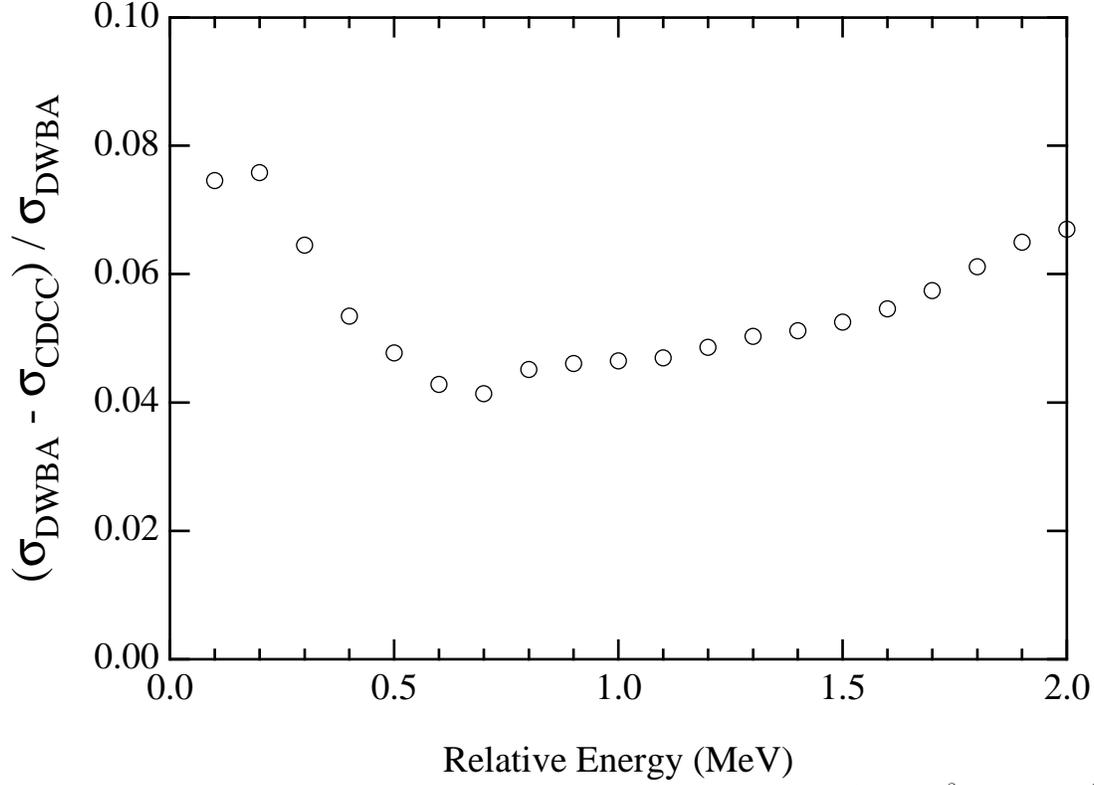} \caption{Difference between the cross
section for Coulomb breakup of 83 MeV/nucleon $^8$B on Pb for $^8$B scattering
angles of 1.77$^\circ$ and less predicted by DWBA (1st-order) and CDCC (all
orders) calculations using the same structure model, expressed as a fraction of
the DWBA prediction. Only Coulomb matrix elements were included in these
calculations. No significant energy dependence of the higher-order
electromagnetic effects is evident.} \label{dwbacomp} \end{figure}

We interpret the required quenching of the $E2$ matrix elements in 1st-order
perturbation theory as a manifestation of higher-order dynamical effects. For a
fixed $E2$ strength, the predicted asymmetry of the longitudinal momentum
distribution is diminished when higher-order effects are considered compared
with 1st-order perturbation theory \cite{esbensen}. In dynamical calculations of
the Coulomb dissociation of $^8$B that include higher-order processes
\cite{esbensen}, the $E1$ strength is essentially unaltered, while the $E2$
strength is reduced with respect to 1st-order perturbation theory calculations.
As such dynamical calculations are difficult and time-consuming, we have
accounted for these effects by quenching the $E2$ matrix elements in the context
of a 1st-order perturbation theory description of the reaction dynamics. The
dynamical calculations include the same physics as do the CDCC calculations
presented here. A comparison between the CDCC calculations and (1st-order) DWBA
calculations using the same structure model indicates that the reduction in $E2$
strength caused by higher-order dynamical effects does not exhibit any
significant relative energy dependence. Fig.\ \ref{dwbacomp} shows the result of
this comparison. Hence the approach we have adopted, namely, scaling the $E2$
matrix elements by the same factor for all relative energies in 1st-order
perturbation theory, is justified.

\begin{figure}\epsfig{file=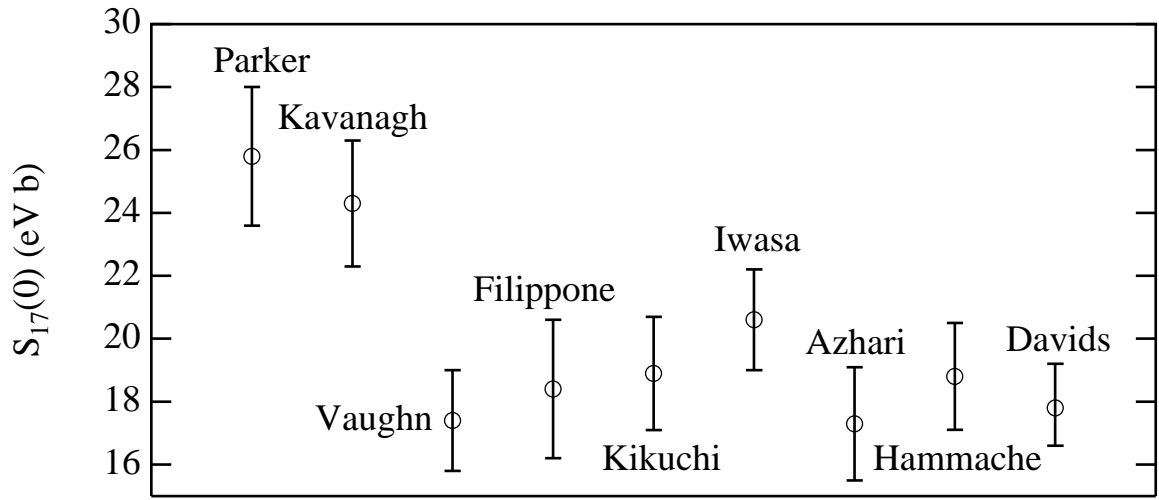} \caption{Inferred zero-energy
astrophysical $S$ factors for the $^{7}$Be(\textit{p},$\gamma$)$^{8}$B reaction
from selected direct and indirect measurements. The data are from ref.s\
\protect\cite{parker,kavanagh69,vaughn,filippone,kikuchi98,iwasa,anc,hammache},
and the present work.} \label{s17comp} \end{figure}

By measuring the Coulomb dissociation cross section at low relative energies
and small scattering angles of the $^8$B center-of-mass, we have ensured that
the contribution of $E2$ transitions is small, and that nuclear diffraction
effects are negligible. Using our inclusive measurement of $^7$Be longitudinal
momentum distributions to determine the relative contributions of the $E2$ and
$E1$ components, we extracted the $E1$ strength at low relative energies from
the exclusive measurement. The value of the astrophysical zero-energy $S$
factor for the $^{7}$Be(\textit{p},$\gamma$)$^{8}$B reaction we infer, 17.8
$^{+ 1.4}_{- 1.2}$ eV b, is in good agreement with other recent measurements,
and with the recommendation of a recent workshop on solar nuclear fusion cross
sections \cite{INT}. Fig.\ \ref{s17comp} and table\ \ref{table4} show the
results of radiative capture, Coulomb breakup, and asymptotic normalization
coefficient determinations of $S_{17}(0)$, along with the results
of this work.

The concordance of our measurement and the other Coulomb breakup measurements
conceals an underlying difference in interpretation. The analyses of references\
\cite{kikuchi98,iwasa} have treated the contributions of $E2$ transitions as
negligible, while our data imply they are not. Since these experiments covered
angular ranges larger than this measurement, they probed smaller impact
parameters where $E2$ transitions are relatively more important. If $E2$
transitions are considered, 1st-order perturbation theory calculations imply
that the astrophysical $S$ factor inferred from the RIKEN Coulomb breakup
measurement should be reduced by 4-15\% \cite{motobayashi00}, and that of the
GSI measurement by 15-20\%. Such a reduction would bring these measurements into
even better agreement with the present work. If we were to analyze our measured
Coulomb breakup cross section between 130 and 400 keV without considering $E2$
transitions, the extracted $E1$ strength would be 5\% greater, and the inferred
value of $S_{17}(0)$ would increase to 18.7 $\pm$ 1.3 eV b. The small $E2$
correction is the result of restricting the angular range covered in this
experiment, making the $E2$ contribution to the breakup cross section comparable
in magnitude to the statistical uncertainty of the measurement.

It appears that the three techniques used to infer $S_{17}(0)$, direct
radiative capture measurements, asymptotic normalization coefficient
determinations, and Coulomb breakup, yield consistent results with different
systematic uncertainties. In light of these facts, we take a weighted average
of these measurements to obtain a recommended value. We include in this average
the recent direct measurements of ref.\ \cite{hammache}, the weighted mean
\cite{anc} of the two published asymptotic normalization coefficient results
\cite{afshin}, and the present Coulomb breakup measurement. Including the
radiative capture measurement of Filippone {\em et al.} \cite{filippone},
which was deemed the only reliable measurement at the 1997 workshop on
solar nuclear fusion cross sections \cite{INT}, makes no difference in the
weighted average. It has been excluded because lack of knowledge about the
target composition prevents accurate correction for the escape of $^8$B
recoils out of the target \cite{hammache}. Similarly, although the data of
ref.s\ \cite{vaughn,hass} are in general consistent with the Hammache {\em
et al.} and Filippone {\em et al.} measurements, the fact that these data
were taken at high energies ($\sim$ 1 MeV), means that one must contend
with substantial extrapolation uncertainties when inferring  $S_{17}(0)$
from them. Since there is a significant dispersion in the inferred values
of $S_{17}(0)$ from such high energy data depending on the $^8$B structure
model used, we have excluded these studies from our weighted average. Among
the direct measurements, that of Hammache {\em et al.} \cite{hammache} is
unique in its careful treatment of both $^8$B backscattering and
theoretical extrapolation errors. We do not include the other Coulomb
breakup measurements \cite{kikuchi98,iwasa} in this average because we lack
sufficient information to precisely correct for the $E2$ component
neglected in the published analyses of these data. The uncertainties in the
considered measurements all contain theoretical contributions, including
extrapolation uncertainties for the radiative capture and Coulomb breakup
measurements. These extrapolation uncertainties are derived from the spread
in the values obtained using different $^8$B structure models for the
extrapolation to zero energy \cite{hammache,jennings}, and vary with the
relative energy ranges considered. The weighted average we obtain is
$\langle S_{17}(0) \rangle $ = 18.0 $\pm$ 0.9 eV b. This value of
$S_{17}(0)$ implies a reduction of the predicted $^8$B solar neutrino flux
of about 5\% from the value used in ref.\ \cite{bahcall}.

\section{Summary}

In summary, we have carried out inclusive measurements of the Coulomb
dissociation of $^8$B on Pb and Ag targets at 44 and 81 MeV/nucleon. Using a
high-resolution, large-acceptance magnetic spectrometer, we measured the
distribution of longitudinal momenta of the emitted $^7$Be fragments. The
longitudinal momentum distributions reveal $E2$ strength in the Coulomb breakup
in the form of an asymmetry produced by $E1$-$E2$ interference. By comparing the
measured longitudinal momentum distributions with 1st-order perturbation theory
calculations, we deduced the effective $E2$ contribution to the Coulomb breakup.
Expressing our result as the ratio of $E2$ and $E1$ $S$ factors at an energy
where previous results have been compiled, we found $S_{E2}/S_{E1}$ = 4.7 $^{+
2.0}_{- 1.3} \times 10^{-4}$ at $E_{rel}$ = 0.6 MeV. This result is at least a
factor of 10 larger than other experimental determinations, but in reasonably
good agreement with theoretical predictions arrived at through several different
methods.

In a separate experiment, we made an exclusive measurement of the Coulomb
dissociation of 83 MeV/nucleon $^8$B on a Pb target using a dipole magnet to
separate the beam from the breakup fragments. Measuring the differential Coulomb
breakup cross section at low relative energies and small $^8$B scattering angles
yielded the astrophysical $S$ factor for the
$^{7}$Be(\textit{p},$\gamma$)$^{8}$B reaction with minimal complications from
$E2$ transitions, higher-order electromagnetic effects, and nuclear-induced
breakup. Interpreting this exclusive measurement in the context of a 1st-order
perturbation theory description of the reaction dynamics and a single-particle
potential model of $^8$B structure, we obtained $S_{17}(0)$ = 17.8 $^{+ 1.4}_{-
1.2}$ eV b. We checked the validity of the perturbative approach through
continuum-discretized coupled channels calculations that assume an essentially
identical model of $^8$B structure. The two reaction theories describe the data
up to relative energies of 2 MeV equally well within the experimental
uncertainties, implying that a slightly modified 1st-order perturbation theory
is adequate for understanding the Coulomb breakup of $^8$B at intermediate beam
energies and small angles.

This measurement agrees well with other recent experimental determinations of
$S_{17}(0)$, and shows that the uncertainties associated with the Coulomb
breakup technique, unwanted multipolarities, higher-order electromagnetic
effects, and nuclear-induced breakup, can be controlled well enough to obtain a
precise value for the $^{7}$Be(\textit{p},$\gamma$)$^{8}$B cross section.
Direct radiative capture measurements, asymptotic normalization coefficient
determinations, and Coulomb breakup measurements yield consistent results for
$S_{17}(0)$, despite their different systematic uncertainties, giving
confidence that this quantity is now well determined. We recommend a weighted
average of measurements using these 3 different techniques, $\langle S_{17}(0)
\rangle $~=~18.0 $\pm$ 0.9 eV b, for use in solar modeling.

\section{Acknowledgments}

This work was supported by the U.S. National Science Foundation under Grant
PHY-95-28844. H.E. was supported by the U.S. Department of Energy, Nuclear
Physics Division, under Contract No. W-31-109-ENG-38. The financial support of
the U.K. Engineering and Physical Sciences Research Council (EPSRC) in the form
of Grant No. GR/M82141, for J.A.T and I.J.T., is gratefully acknowledged.

\begin{table} \caption{Total number of $^{8}$B nuclei on target} \begin{center}
\begin{tabular}{|c|c|c|} Target & Beam Energy & $^8$B on Target \\ &
(MeV/nucleon) & (10$^{6}$) \\ \hline Ag & 44 &   360 \\ Ag & 81 & 1 070\\ Pb &
44 & 840 \\ Pb & 81 & 2 980 \label{table1} \end{tabular} \end{center}
\end{table}

\begin{table} \caption{Integrated Coulomb dissociation cross sections}
\begin{center} \begin{tabular}{|c|c|c|c|}Target & Beam Energy & $^7$Be angle cut
($^\circ$) & $\sigma$ (mb) \\ & (MeV/nucleon) & & \\ \hline Ag & 44 & 1.5 & 61
(7) \\ Ag & & 2.0 & 97 (10) \\ Ag & & 2.5 & 140 (15) \\ Pb & & 1.5 & 68 (7) \\
Pb & & 2.4 & 156 (16) \\ Pb & & 3.5 & 252 (25) \\ Ag & 81 & 1.25 & 67 (7) \\ Pb
& & 1.5 & 130 (8) \\ Pb & & 2.0 & 201 (13) \\ Pb & & 2.5 & 266 (17)
\label{table2} \end{tabular} \end{center} \end{table}

\begin{table} \caption{Comparison of theoretical $E2$ strength predictions with
present results} \begin{center} \begin{tabular}{|c|c|c|c|} Author & Reference &
Method & $S_{E2}/S_{E1}$ (0.6 MeV) \\ \hline Esbensen and Bertsch &
\cite{esbensen} & potential model & 9.5 $\times10^{-4}$ \\ Typel {\em et al.} &
\cite{typel} & potential model & 8.0 $\times10^{-4}$ \\ Bennaceur {\em et al.} &
\cite{bennaceur} & SMEC & 7.72 $\times10^{-4}$ \\ Descouvement and Baye &
\cite{descouvement} & cluster model & 6.2 $\times10^{-4}$ \\ Barker &
\cite{barker} & R-Matrix & 8.7 $\times10^{-4}$ \\ Davids {\em et al.} & this
work & experiment & $4.7$ $^{+ 2.0}_{- 1.3}$ $\times10^{-4}$ \label{table3}
\end{tabular} \end{center} \end{table}

\begin{table} \caption{Recent $S_{17}(0)$ determinations} \begin{center}
\begin{tabular}{|c|c|c|c|} Author & Reference & Method & $S_{17}(0)$ (eV b) \\
\hline Filippone {\em et al.} reanalysis & \cite{filippone,hammache} &
radiative capture & 18.4 $\pm$ 2.2 \\ Hammache {\em et al.} & \cite{hammache} &
radiative capture & 18.8 $\pm$ 1.7 \\ Kikuchi {\em et al.} & \cite{kikuchi98} &
Coulomb breakup & 18.9 $\pm$ 1.8 \\ Iwasa {\em et al.} & \cite{iwasa} & Coulomb
breakup & 20.6 $\pm$ 1.0 $\pm$ 1.0 \\ Azhari {\em et al.} & \cite{anc,afshin} &
transfer reaction & 17.3 $\pm$ 1.8 \\ Davids {\em et al.} & this work & Coulomb
breakup & 17.8 $^{+ 1.4}_{- 1.2}$ \label{table4} \end{tabular} \end{center}
\end{table}

\end{document}